\documentclass[pra,twocolumn,showpacs,aps,floatfix,amsmath,amssymb,amsfonts]{revtex4-1}

\usepackage{amsmath}
\usepackage{graphicx}
\usepackage{color}

\newcommand{\mbf}{\mathbf}
\newcommand{\mrm}{\mathrm}
\newcommand\ii{{\rm i}}
\newcommand{\beq}{\begin{equation}}
\newcommand{\eeq}{\end{equation}}

\begin{document}

\title{Properties of strongly dipolar Bose gases beyond the Born approximation}
\author{Rafa{\l} O{\l}dziejewski$^1$}
\author{Krzysztof Jachymski$^2$}
\affiliation{
$^1$ Center for Theoretical Physics, Polish
Academy of Sciences, Al. Lotnik\'{o}w 32/46,
02-668 Warsaw, Poland\\
$^2$ Institute for Theoretical Physics III, University of Stuttgart, Pfaffenwaldring 57, 70550 Stuttgart, Germany
}
\date{\today}

\begin{abstract}
Strongly dipolar Bose gases can form liquid droplets stabilized by quantum fluctuations. In theoretical description of this phenomenon, low energy scattering amplitude is utilized as an effective potential. We show that for magnetic atoms corrections with respect to Born approximation arise, and derive modified pseudopotential using realistic interaction model. We discuss the resulting changes in collective mode frequencies and droplet stability diagram. Our results are relevant for recent experiments with erbium and dysprosium atoms.
\end{abstract}

\maketitle

\section{Introduction}
Ultracold gases with dipolar interactions have been the subject of intense experimental and theoretical studies over the past few years~\cite{Lahaye:2009}.
Strong dipole-dipole interactions can be realized by using atoms with large magnetic moments such as chromium, erbium and dysprosium~\cite{Lahaye:2007,Lu:2011,Aikawa:2012} or polar molecules with electric dipole moment induced by static external field~\cite{Ni:2008,Park2015}.
Compared to systems with contact interactions, dipolar gases exhibit much richer physics both in the few-body~\cite{Wang2015} and many-body~\cite{baranov2012condensed} domains. 
Even in the limit of weak interactions, their properties are highly affected by the long-range and anisotropic nature of the dipole-dipole interaction term.
In particular, for sufficiently strong dipole-dipole interactions Bogoliubov excitation spectrum of the Bose gas can contain unstable modes which drives the collapse of the condensate, observed experimentally e.g. with chromium~\cite{Lahaye2008}.

Recent experiments performed using atoms with high magnetic moments revealed the existence of new intriguing effects. When the condensate is quenched into the regime unstable from the mean field point of view, instead of collapsing the gas can form a spatially ordered structure of stable droplets with high density~\cite{Barbut2016,Wachtler2016,Kadau2016}. Stabilization of the gas in form of the droplets is due to quantum fluctuations which have crucially important contribution near the instability. These quantum droplets turn out to be stable even without the presence of an external trap, with lifetime limited only by three-body losses~\cite{Wachtler2016a,Chomaz2016,Baillie2016,Schmitt2016}.

In the regime of weak interactions the mean field description of dipolar Bose gas relies on modified Gross-Pitaevskii (GP) equation which acquires an additional nonlocal term coming from dipole-dipole interactions~\cite{Yi2001}.
Description of the droplet phase requires including beyond mean field Lee-Huang-Yang (LHY) correction which provides an additional effectively repulsive term preventing the gas from collapsing. For dipolar gases the LHY correction is strongly enhanced compared to contact interactions~\cite{Schutzhold2006,Lima2011,Lima2012}.

In the mean field theory for interacting bosons, the effective potential is given by the low energy scattering amplitude~\cite{Beliaev1958,Hugenholtz1959}. In contrast to contact interactions, for dipolar bosons it is required to include many partial waves in the scattering calculation. For weak dipole moments and away from scattering resonances Born approximation describes the scattering amplitude with good accuracy~\cite{Yi2001,Kanjilal2008,Ticknor2008,Bohn:2009,Zhang2014,Bohn2014}. However, for lanthanide atoms the validity of Born approximation can be limited as the magnetic dipole-dipole interaction becomes too strong. Furthermore, the interaction potential at short range has a complex anisotropic structure, resulting in additional coupling between partial waves. As a result, lanthanide atoms exhibit very dense structures of Feshbach resonances~\cite{Chin:2008,Petrov:2012,Baumann:2014,Frisch2014,Maier2015}. In the vicinity of a resonance Born approximation is not valid at all and one has to use numerically calculated scattering amplitude. This can modify the many-body properties of the dipolar gas~\cite{Wang2008,Huang2010}.

While the droplet model based on extended GP equation within Born and local density approximations has so far successfully described the droplet formation and their basic properties, it is worthwhile to study more elaborate models for deeper understanding of the droplet phase.
In particular, finite temperature effects can affect the size and lifetime of the droplets~\cite{Boudjemaa2016}.
Effects of the trapping potential can also be important especially when the characteristic trap lengths become comparable not only to the condensate healing length, but also to the length scales characteristic for the scattering.
Finally, strong dipolar interactions and dense resonances limit the usefulness of Born approximation for the scattering amplitude.
In this work, we concentrate on the latter point and study how the properties of dipolar Bose gas are affected by using realistic interaction potentials.
We find that the role of dipolar interactions is enhanced away from resonances and strongly suppressed in their vicinity. This has visible consequences on the many-body properties of the gas, most notably on the critical atom number needed for formation of a stable droplet.

The paper is structured as follows. In Section II we compute the scattering amplitude for a pair of magnetic atoms using a realistic interaction potential and discuss the validity of Born approximation. In Section III we briefly review the mean field theory of dipolar quantum gases and the role of beyond mean field corrections, and discuss the stability diagram of a trapped gas. These results are then used in Section IV to analyze the experimentally relevant cases of Dy atoms and weakly bound Dy$_2$ molecules. Conclusions are drawn in Section V.

\section{Scattering amplitude}
\label{sec:scater}

Providing a reasonable estimate for the low energy scattering amplitude of lanthanide atoms is a demanding task because of the complex interaction potential~\cite{Petrov:2012}. The dominating terms are isotropic van der Waals attraction $V_{\rm vdW}=-C_6/r^6$ and anisotropic dipole-dipole interaction $V_{\rm dd}=-C_3 P_2(\cos\theta)/r^3$. Other terms such as quadrupole-quadrupole and anisotropic van der Waals interaction introduce additional couplings between different partial waves and hyperfine channels. This leads to occurrence of extremely densely spaced Feshbach resonances with high partial wave character~\cite{Frisch2014}. In our model we simulate this behavior by varying the short-range boundary conditions to mimic a Feshbach resonance and we focus on the interplay of the van der Waals and dipolar interaction.

\begin{figure}
\centering
\includegraphics[width=0.45\textwidth]{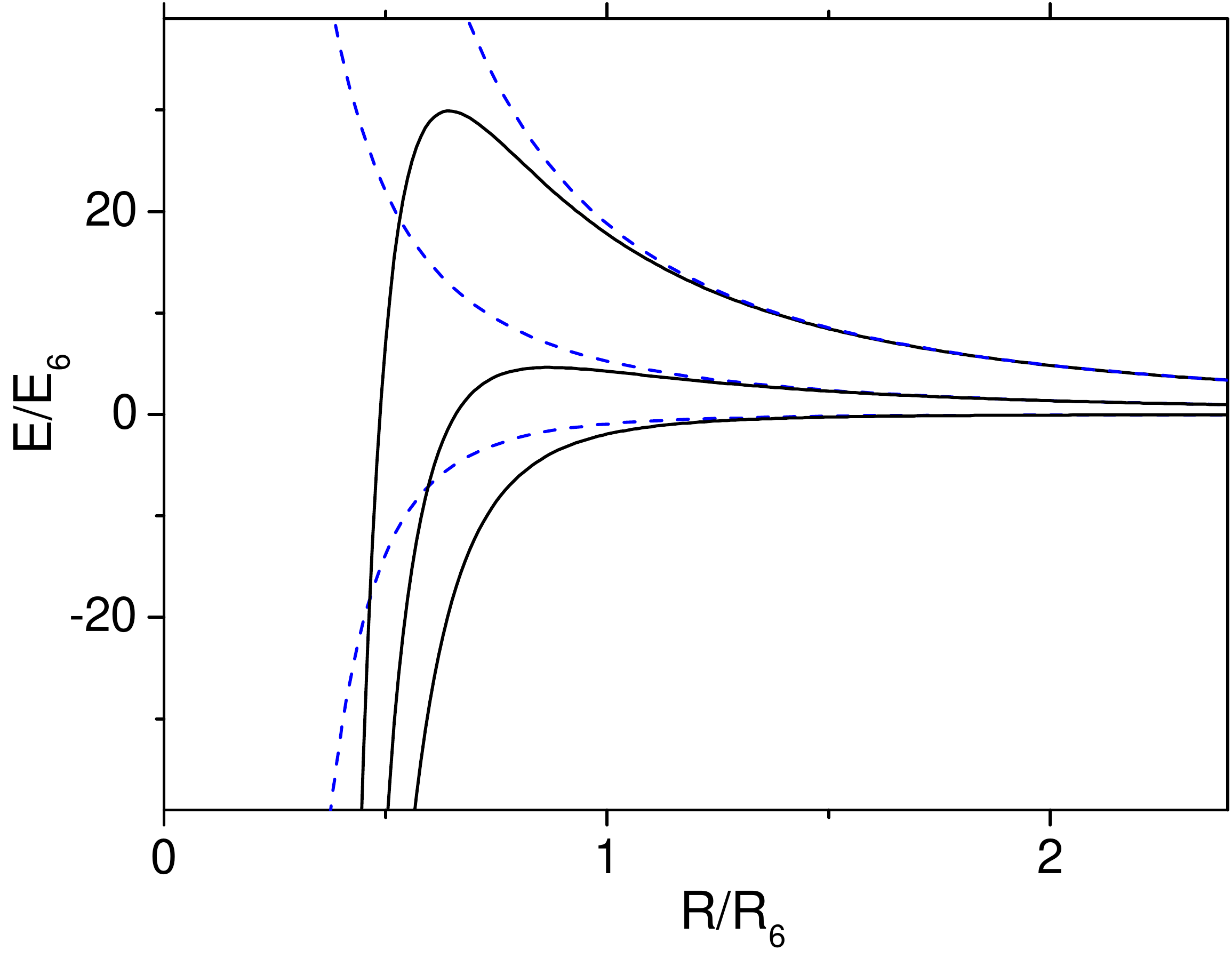}
\caption{\label{fig:adpots}Three lowest adiabatic potential curves with $m=0$ obtained for realistic dysprosium long range potential (black lines) and for the case where van der Waals interaction is replaced by a hard wall set at short distance. The length unit here $R_6=\left(2\mu C_6/\hbar^2\right)^{1/4}$, while the energy is given in units of $E_6=\hbar^2/\left(2\mu R_6 ^2\right)$.}
\end{figure}

A commonly used treatment of ultracold dipolar scattering involves using a simplified potential with the dipolar term and a hard wall placed at some distance $r_0$ to set the scattering length~\cite{Kanjilal2008,Bohn:2009,Oldziejewski2016} and utilizing Born approximation to obtain the scattering amplitude. However, for magnetic atoms the length scales associated with different terms in the potential are comparable, with strong coupling between partial waves. Born approximation can be expected to break down in this case, especially in the vicinity of scattering resonances. The prominent role of the van der Waals potential can be seen e.g. by studying the adiabatic potential curves obtained by diagonalizing the angular part of the Hamiltonian in partial wave basis. Figure~\ref{fig:adpots} shows a comparison between realistic adiabatic potentials and the hard wall approximation for the case of dysprosium. It can be seen that including van der Waals interaction leads to vastly different shape and height of the centrifugal barrier. As for higher partial waves the position of the barrier moves towards the origin, the modification is even stronger in high partial wave channels.

Let us now discuss the relevant length and energy scales. In this work we will focus on strongly magnetic dysprosium atoms, for which the isotropic van der Waals coefficient takes the value $C_6=2270$ in atomic units~\cite{Lepers2016} and the dipole moment is $9.93\mu_B$ where $\mu_B$ is the Bohr magneton. One can then define characteristic van der Waals length as $R_6=\left(2\mu C_6/\hbar^2\right)^{1/4}$, which for dysprosium gives $161\,a_0$ with $a_0$ being the Bohr radius. Dipolar interactions can be characterized in a similar way by $R_{\rm dd}=2\mu d^2/ 3\hbar^2$, for Dy atoms giving $R_{\rm dd}=131\,a_0$. 
We will also analyze the case of weakly bound Dy$_2$ dimers for which we take $C_6=4C_6^{\rm Dy}$ and the dipole moment is assumed to be two times larger than for Dy. This kind of molecules can be produced by making use of broad universal Feshbach resonances~\cite{Maier2015}. Weakly bound erbium molecules have been recently created and characterized~\cite{Frisch:2015}. The characteristic length scales for this case are $R_6=271.5\,a_0$ and $R_{\rm dd}=1049\,a_0$. In atomic units the dipole moments we are considering here are equivalent to $0.1-0.2\,$D. These values are easily achievable also with polar molecules placed in external electric field.

To solve the low energy scattering problem with realistic potential, we first set the boundary conditions at very short range where the van der Waals interaction dominates and coupling between partial waves is negligible. The phase of the short range wave function would then determine the scattering length for pure van der Waals interactions. Then we propagate the solution to large distances using Numerov method with collision energy set to be smaller than all characteristic energies of the problem. Dipolar interactions require including a huge number of partial waves and propagating the wave function to very large distances to obtain convergent results~\cite{Ticknor2008,Bohn:2009}. Similarly to changing the position of the hard wall, we can manipulate the short range phase to set the scattering length and simulate a Feshbach resonance (note that in contrast to contact interactions, the actual scattering length differs from the position of the hard wall as it is renormalized by dipolar interaction). The long-range nature of dipolar term which behaves as $r^{-3}$ results in increased density of near threshold bound states, so usually several resonances appear when changing the short range phase. Additional couplings can easily be included within this method.

We fit the wave function at large distances to asymptotic scattering solutions to extract the scattering amplitude
\beq
f(\mathbf{k},\mathbf{k^\prime})=4\pi\sum_{\ell m \ell^\prime m^\prime}{Y_{\ell m}^\star (\hat{k})Y_{\ell^\prime m^\prime}(\hat{k}^\prime) t_{\ell m}^{\ell^\prime m^\prime}(k)}.
\label{eq:fkk}
\eeq
where $t_{\ell m}^{\ell^\prime m^\prime}$ is the reduced T-matrix elements and $Y$ denote standard spherical harmonics. For bosons the summation runs only over even values of $\ell$, $\ell^\prime$.
For dipolar interactions Born approximation predicts~\cite{Yi2001,Bohn2014} 
\beq
\left(t_B\right)_{\ell m}^{\ell^\prime m^\prime} = -\frac{6\pi \left<Y_{\ell, m}\right|Y_{2,0}\left|Y_{\ell^\prime, m^\prime}\right> R_{\rm dd}}{((\ell-\ell^\prime)^2-1)(\ell+\ell^\prime)(\ell+\ell^\prime+2)}
\eeq
for $\ell+\ell^\prime>0$. For the $t_{00}^{00}$ element one has to insert the value of the $s$-wave scattering length $a$. The total scattering amplitude within the Born approximation is then~\cite{Bohn:2009,Bohn2014}
\beq
f_B(\theta)=-a-a_{\rm dd}^{\rm Born}P_2(\cos\theta),
\label{eq:fBorn}
\eeq
where $\theta$ is the angle between the momentum transfer vector and the dipole axis and $a_{\rm dd}^{\rm Born}=R_{\rm dd}$. The relative strength of the dipolar interaction in the Born approximation can be characterized by $\epsilon_{\rm dd}^{\rm Born}=a_{\rm dd}^{\rm Born}/a$.

\begin{figure}
\centering
\includegraphics[width=0.5\textwidth]{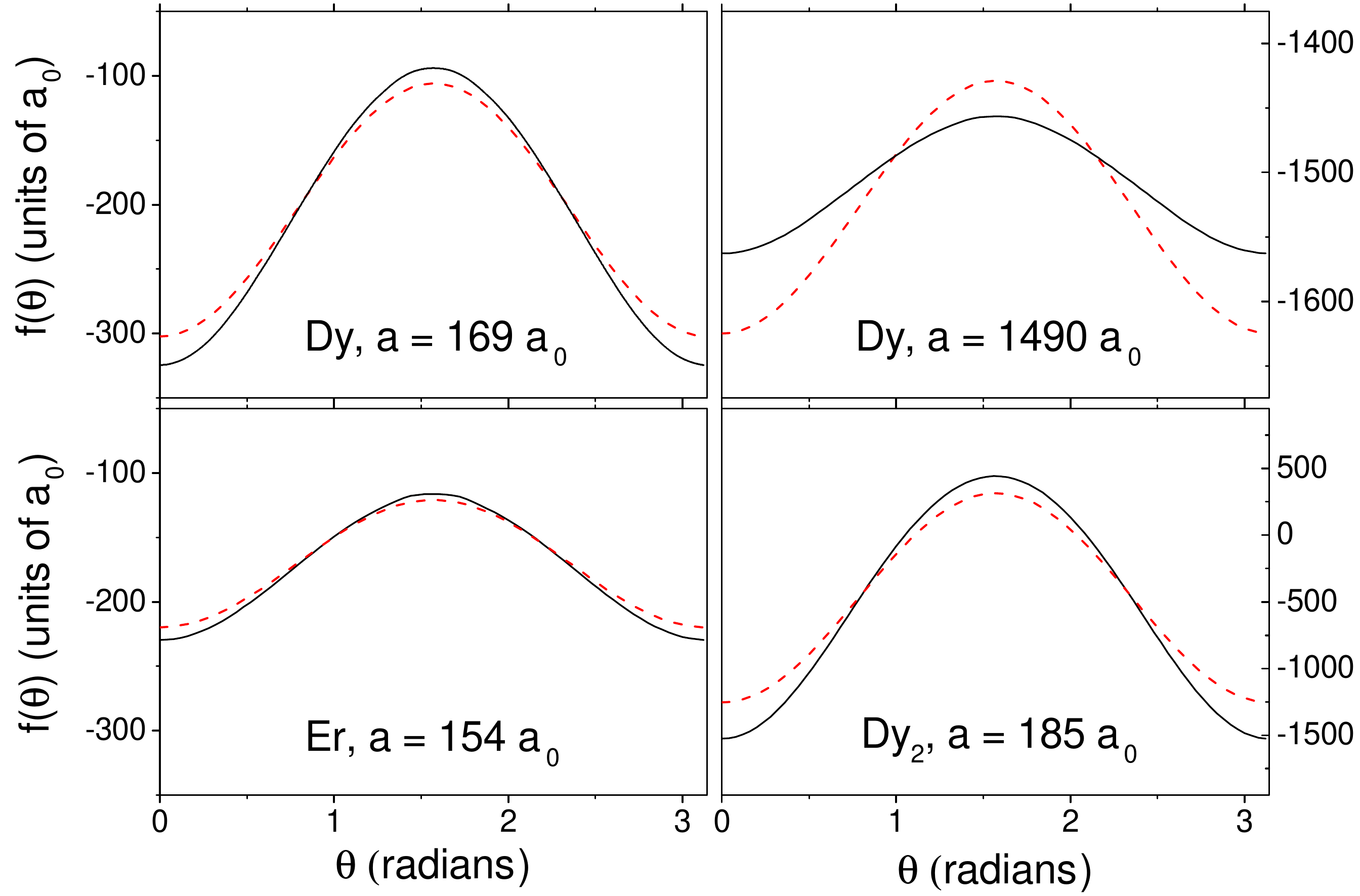}
\caption{\label{fig:fDy}Scattering amplitude $f(\theta)$ for dysprosium atoms at scattering length $a=169\,a_0$ (upper left) and $a=1490\,a_0$ (upper right), for erbium with $a=154\,a_0$ (lower left) and for Dy$_2$ dimers with $a=185\,a_0$ (lower right). Straight black lines come from the numerical calculation, while red dashed lines show the Born approximation predictions. Note that different scale was used for Dy$_2$, where the amplitude spans a wider range due to much larger $a_{\rm dd}$ coefficient.}
\end{figure}

\begin{figure}
\centering
\includegraphics[width=0.5\textwidth]{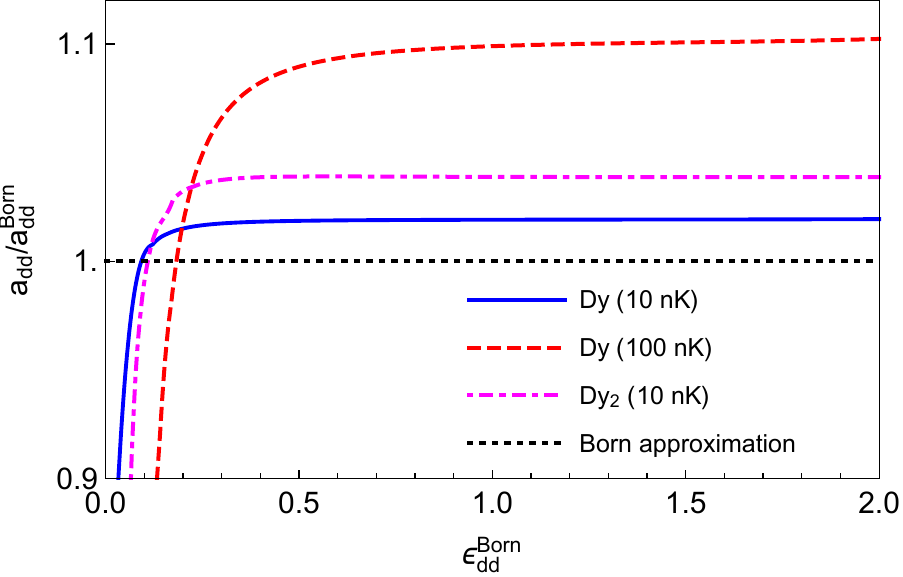}
\caption{\label{fig:add}Effective dipolar length as a function of $\epsilon_{\rm dd}^{\rm Born}$ for dysprosium at 10nK (blue straight line) and at 100nK (red dashed line) and Dy$_2$ (dot-dashed magenta line). Dotted black line shows the Born approximation result $a_{\rm dd}/a_{\rm dd}^{\rm Born}=1$.}
\end{figure}

\begin{figure}
\centering
\includegraphics[width=0.3\textwidth]{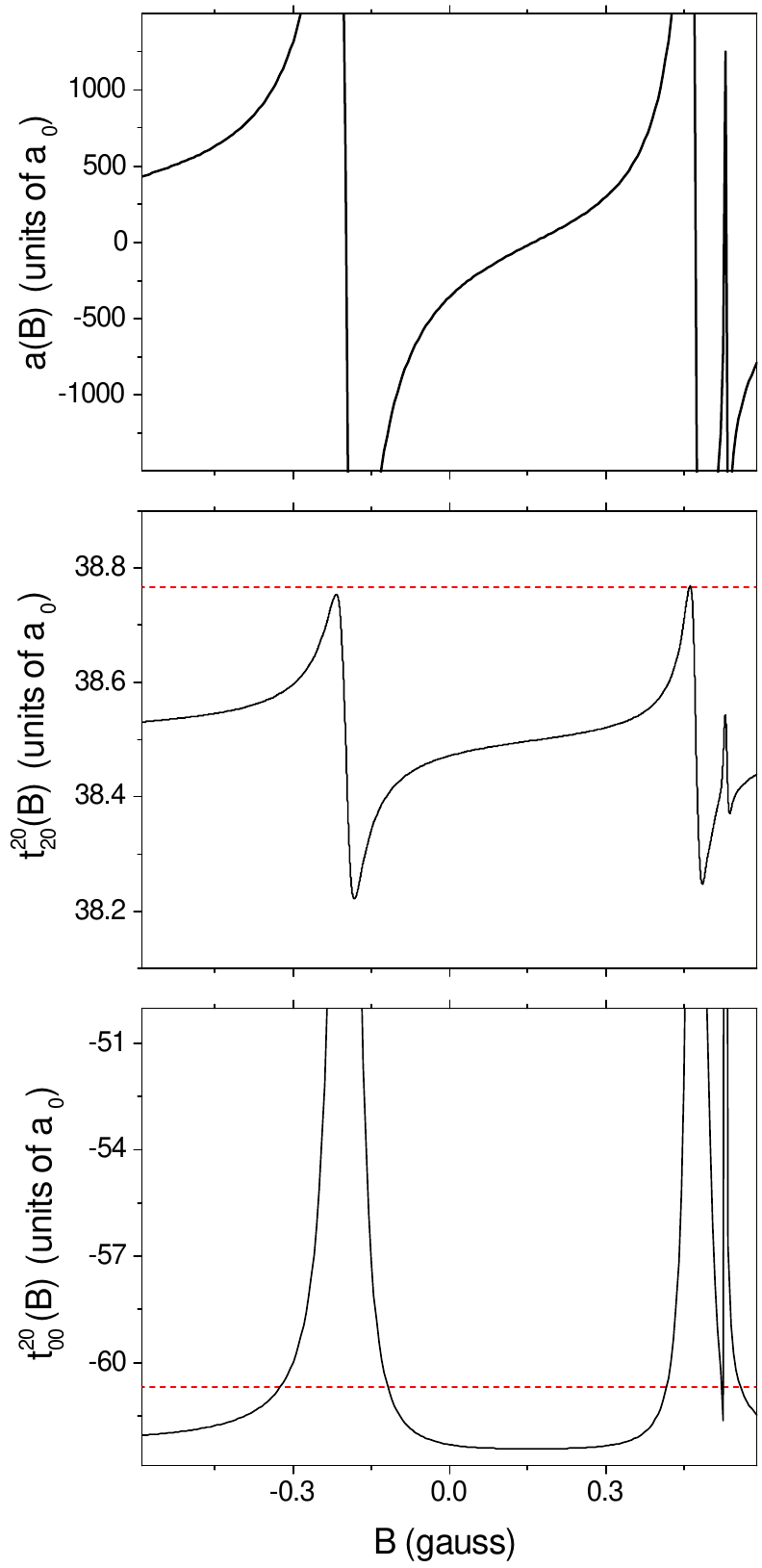}
\caption{\label{fig:res}Exemplary magnetic field dependencies of the lowest $t$ matrix elements near a Feshbach resonance in dysprosium. Black lines show the numerical results, while red dashed are the Born approximation results.}
\end{figure}

Figure~\ref{fig:fDy} shows exemplary scattering amplitudes for Dy atoms away from resonance and close to it, as well as for Er atoms and Dy$_2$ dimers away from resonance compared with Born approximation predictions. Surprisingly, in all the cases we investigated the numerically calculated scattering amplitude could be well described with the same form as in Eq.~\eqref{eq:fBorn}, but with modified $a_{\rm dd}$ parameter (from now on we will refer to the standard dipolar length as $a_{\rm dd}^{\rm Born}$ as it remains unchanged within Born approximation). This result simplifies a lot the analysis of many-body effects, since all the methods developed for dipolar bosons within Born approximation can be readily used. The next observation is that away from resonances, where scattering length is of the order of $a_{\rm dd}^{\rm Born}$, the effective $a_{\rm dd}$ turns out to be slightly larger than the actual one due to corrections arising from all partial waves. The modification turns out to be strongly energy-dependent. For Dy at 10nK collision energy the enhancement of effective dipolar length is of the order of $2\%$ and at 100nK about 10\%, while for Dy$_2$ it is about $4\%$ at 10nK. Increasing the value of the scattering length leads to a significant deviation of low $(\ell,m)$ $t$ matrix elements from Born approximation (see Fig.~~\ref{fig:res}) and results in much smaller effective $a_{\rm dd}$. Figure~\ref{fig:add} shows the dependence of the effective dipolar length on the dipolar interaction strength $\epsilon_{\rm dd}$. We see that the impact of the resonance depends on the dipole moment. For low dipole moments the dependence of $a_{\rm dd}$ on the scattering length is changed only very close to the resonance, while for highly dipolar particles the resonance has visible impact in a larger range of scattering lengths. As a result, the actual relative strength of the dipolar interaction defined as $\epsilon_{\rm dd}=a_{\rm dd}/a$ differs from $\epsilon_{\rm dd}^{\rm Born}$, being larger away from resonances and smaller in the vicinity of a resonance.

Figure~\ref{fig:res} shows an example of the dependence of the lowest $t$ matrix elements near an exemplary Feshbach resonance. Scattering length has more than one pole here due to increased density of near threshold bound states resulting from dipolar interaction. At each resonance, $t^{20}_{20}$ also diverges (whis is not clearly visible on the figure due to too low resolution of the data points and finite collision energy which suppresses the divergence) and $t^{20}_{00}$ vanishes. Within our model, channels with $m\neq 0$ are not affected by the resonance. It is possible to include additional couplings that modify scattering in higher partial wave channels. However, the phase shifts in these channels would still be dominated by universal dipolar scattering and the expected effects would be small as long as the collision energy is low enough.

\section{Effective mean field theory for dipolar bosons}
Having calculated the scattering amplitude for realistic interaction potential, we will now use these results in the effective many-body theory. 
The system under consideration consists of $N$ magnetic (or electric) dipolar bosons of mass m and dipole moment $d$ oriented along the $z$ direction by an external magnetic field trapped in an external potential $V_{\mrm ext}({\mbf r})=\frac{m}{2}\left(\omega_x^2 x^2+\omega_y^2 y^2+\omega_z^2 z^2\right)$ with $\omega_{x,y,z}$ the trapping frequencies. The second quantized Hamiltonian reads

\begin{eqnarray}
 H&=& \int\! d{\mbf r}\,\hat\Psi^{\dag}({\mbf r})
\left[-\frac{\hbar^2}{2m}\nabla^2+V_{\mrm ext}({\mbf r})\right]\hat\Psi({\mbf r}) \nonumber\\
&\ &+\frac{1}{2}\int\! d{\mbf r}\int\! d{\mbf r'}
\hat\Psi^{\dag}({\mbf r})\hat\Psi^{\dag}({\mbf r'})
V({\mbf r}-{\mbf r'})\hat\Psi({\mbf r'})\hat\Psi({\mbf r}),
\label{h}
\end {eqnarray}

where $\hat\Psi({\mbf r})$ and $\hat\Psi^{\dag}({\mbf r})$ are bosonic annihilation and creation field operators respectively and $V({\mbf r}-{\mbf r'})$ is the interaction potential between the atoms. 

The standard way to obtain the mean field Gross-Pitaevskii equation is to replace the field operators by a complex-valued function $\psi(\mbf r,t)=\left\langle\hat\Psi(\mbf r,t)\right\rangle$. Here we also have to include the LHY correction to the mean field equation of state. 
Furthermore, the true interaction potential between the atoms $V({\mbf r}-{\mbf r}')$ is replaced by the effective potential $V_{\rm eff}({\mbf r}-{\mbf r}')$ given by the low energy scattering amplitude~\cite{Beliaev1958,Hugenholtz1959} obtained for the full potential. A general expression for an anisotropic effective potential can be given as~\cite{Yi2001}
\begin{eqnarray}
V_{\rm eff}({\mbf r})=\frac{4\pi \hbar^2  a}{m}\delta({\mbf r}) +\sum_{l>0, m}
\alpha_{lm} {Y_{lm}(\hat r)\over |{\mbf r}|^{3}}, \label{veff}
\end{eqnarray}
where the first term describes the $s$-wave contribution, $Y_{lm} $ stays for standard spherical harmonic and $\alpha_{lm}$ are determined from the scattering amplitude~\cite{Yi2001}. According to the results from Section II in our case the effective potential may be written as the well-known formula consisting of a sum of the contact and dipole-dipole interactions:
 \begin{eqnarray}
V_{\rm eff}({\mbf r}-{\mbf r}')=\frac{4\pi \hbar^2  a}{m}\delta({\mbf r})-\frac{4\pi \hbar^2  a_{\rm dd}}{m}{2P_{2}(cos{\theta})\over |{\mbf r}-{\mbf r}'|^{3}}, \label{nasz}
\end{eqnarray}
where $P_{2}(\cos{\theta})$ is the second Legendre polynomial. We remind that $a_{\rm dd}$ is an effective parameter which stems from the scattering amplitude.

The extended GP equation within the local density approximation for slowly varying spatial profile of $\psi(\mbf r,t)$ takes the form~\cite{Wachtler2016}
\begin{eqnarray}
\ii\hbar \frac{\partial}{\partial t}\psi({\mbf r},t) &=&  \left [ -\frac{\hbar^2}{2m}\nabla^2+V_{\rm ext}({\mbf r})   \right\delimiter 0 \nonumber \\ 
 && \left\delimiter 0  +N\int d^3r'V_{\rm eff}({\mbf r}-{\mbf r}')|\psi({\mbf r},t)|^2\right\delimiter 0 \nonumber \\ 
 && \left \delimiter 0 + g_{\rm QF}N^{3/2}|\psi({\mbf r},t)|^{3}  \right ]      \psi({\mbf r},t), 
 \label{GPE}
\end{eqnarray} 
with $\int d^3r |\psi({\mbf r},t)|^2=1$. The last line of Eq.~\eqref{GPE} refers to the LHY correction to the equation of state. It can be intuitively understood as the zero-point motion of the Bogolibuov excitations. Primarily it was derived for a homogeneous gas with the contact interaction. However, it can be generalized to the case of dipolar interactions as well~\cite{Lima2011,Lima2012}. It turns out that the LHY contribution scales as $N^{3/2}$ for both contact and dipolar case. The magnitude of the quantum fluctuations can be described by the coefficient $g_{\mrm QF}$. For the dipolar interaction $g_{\mrm QF}=\frac{32}{3\sqrt{\pi}}g\sqrt{a^3}f(\epsilon_{\rm dd})$, where $f(\epsilon_{\rm dd})=\frac{1}{2}\int_0^{\pi} d\theta \sin{\theta}(1+\epsilon_{\rm dd}(3\cos^2\theta-1)) ^{5/2}$. The relative strength of the dipolar interaction is given by $\epsilon_{\rm dd}$. As we can see the LHY correction has a universal character, because it depends only on the two-body scattering length and the relative strength of the dipolar interaction. The same formula, only with effective $a_{\rm dd}$, can be used in our model.

A quite convenient way to examine the basic properties of strongly dipolar bosons such as the ground state or elementary excitations is to use a simple variational Gaussian ansatz for the wave function. Even though the system is in the Thomas-Fermi regime where the shape of the atomic cloud is far from being Gaussian, it has been shown that results obtained from this simple method are in considerable agreement with those obtained by solving numerically the Gross-Pitaevskii equation~\cite{Baillie2016,Wachtler2016a}. In the Gaussian model we take~\cite{Wachtler2016a}:

\begin{equation}
 \psi(x,y,z)=\frac{1}{\pi^{\frac{3}{4}}\left(\sigma_x \sigma_y \sigma_z\right)^{\frac{1}{2}}} \prod_{\alpha=x,y,z} {\rm e}^{-\frac{\alpha^2}{2\sigma_\alpha^2}+i\alpha^2 \eta_{\alpha}}, \label{GaussianAnsatz}
\end{equation}
where the variational parameters are the widths $\sigma_\alpha(t)$ in the $\alpha=x,y,z$ direction, and $\eta_\alpha(t)$, which determines the phase curvature along $\alpha$. We follow the notation proposed by W\"{a}chtler and Santos~\citep{Wachtler2016a,Chomaz2016}.  By straightforward calculation one can derive the energy functional corresponding to Eq.~\eqref{h} and then determine the following Euler-Lagrange equations:

\begin{equation}
\eta_\alpha=\frac{m}{2\hbar \sigma_\alpha} \frac{d\sigma_\alpha}{dt}
\end{equation}
and 
\begin{equation}
\frac{d^2 \nu_\alpha}{d \tau^2}=-\frac{\partial}{\partial \nu_\alpha}U(\nu_x,\nu_y,\nu_z),
\end{equation}
where in the second equation dimensionless units $\tau=\tilde{\omega}t $, $\sigma_\alpha=\tilde{l}\nu_\alpha$, $\tilde{l}=\sqrt{\hbar/m \tilde{\omega}}$ are introduced, with $\tilde\omega=(\prod\omega_\alpha)^{1/3}$. The potential $U(\nu_x,\nu_y,\nu_z)$ for an arbitrarily chosen anisotropic potential expressed as in Eq. \eqref{veff} may be written as

\begin{eqnarray}
 U&=& \frac{1}{2}\sum_{\alpha}\left [ \nu_\alpha^{-2}+\left ( \frac{\omega_\alpha}{\tilde\omega} \right )^2\nu_\alpha^2 \right ]+\frac{2}{3}\frac{PAN^{3/2}}{\left(\prod_\alpha \nu_\alpha \right)^{\frac{3}{2}}}\nonumber \\
 &+&  \frac{PN}{\prod_\alpha \nu_\alpha}
 -\frac{1}{2}\int\int d^3r\,d^3r'\times
\nonumber \\ &\times & \sum_{l>0, m}\alpha_{lm} {Y_{lm}({\hat{\mbf R}})\over |{\mbf R}|^{3}}|\psi({\mbf r},t)|^2\psi({\mbf r'},t)|^2  ,
 \label{Energy_vi}
\end{eqnarray} 
where ${\mbf R}={\mbf r}-{\mbf r'}$ and the dimensionless constants $P=\sqrt{\frac{2}{\pi}}\frac{a}{\tilde{l}}$ and $A\sim \left ( \frac{a}{\tilde{l}} \right )^{3/2}$, which describes the magnitude of the LHY correction for a given effective potential $V_{\mrm eff}$. For the effective potential described by Eq.~\eqref{nasz} the expression for $U(\nu_x,\nu_y,\nu_z)$ takes exactly the same form as in Ref.~\cite{Chomaz2016}, only with effective $a_{\rm dd}$.

In order to describe the ground state properties one has to find the minimum of $U$ that corresponds to the equilibrium widths $\nu_\alpha^{eq}$. Then the lowest excitations are fully determined by the Hessian matrix at the minimum of $U$. For spherical or cylindrical harmonic traps the three lowest excitations are characterized by the well-known 3D monopole, 3D quadrupolar and 2D quadrupolar modes~\citep{Wachtler2016a,Chomaz2016}.

To analyze stability of a self-bound droplet solution it is convenient to calculate the release energy $E_{\mrm R}$, which is defined as the energy of the system when subtracting the energy related with the confinement, at the equilibrium widths corresponding to a droplet state~\citep{Chomaz2016}. Whenever $E_{\mrm R}<0$ holds a self-bound solution is reached.

\section{Results}
Our goal is now to compare recent results concerning strongly interacting dipolar bosons in Ref.~\citep{Wachtler2016a,Chomaz2016,Baillie2016} with our calculations based on the effective potential  obtained from the realistic scattering amplitude described in Sec. II. In order to simplify our presentation we will describe the results as a function of $\epsilon_{\rm dd}^{\rm Born}=a_{\rm dd}^{\rm Born}/a$ and $\epsilon_{\rm dd}=a_{\rm dd}/a$. The relation between both relative dipolar strengths depends on the scattering length, as shown by~Fig.~\ref{fig:add}. We note that within Born approximation the theoretical predictions are completely universal, as the properties of the system depend only on the value of $\epsilon_{\rm dd}^{\rm Born}$. In contrast, in our treatment the dependence of $\epsilon_{\rm dd}$ on the scattering length is system specific.

\subsection{Droplet stability}
Theoretical calculations~\cite{Wachtler2016a,Chomaz2016,Baillie2016} predict the existence of stable self-bound droplets stabilized by quantum fluctuations (the LHY correction) in the region where mean-field solution predicts a collapse. In the recent experiment~\cite{Schmitt2016} the self-bound droplet has indeed been observed. However, the phase boundary (the critical atom number $N_{\rm crit}$ as a function of the scattering length) was not fully reproduced by theory. The discrepancy was explained by proposing a shift in the background scattering length from $92$ to $62.5\,a_0$. As the measurement of the critical atom number is very sensitive to the interaction parameters, the correction to the effective interaction can also significantly alter the phase diagram.

\begin{figure}[h]
\centering
\includegraphics[width=0.5\textwidth]{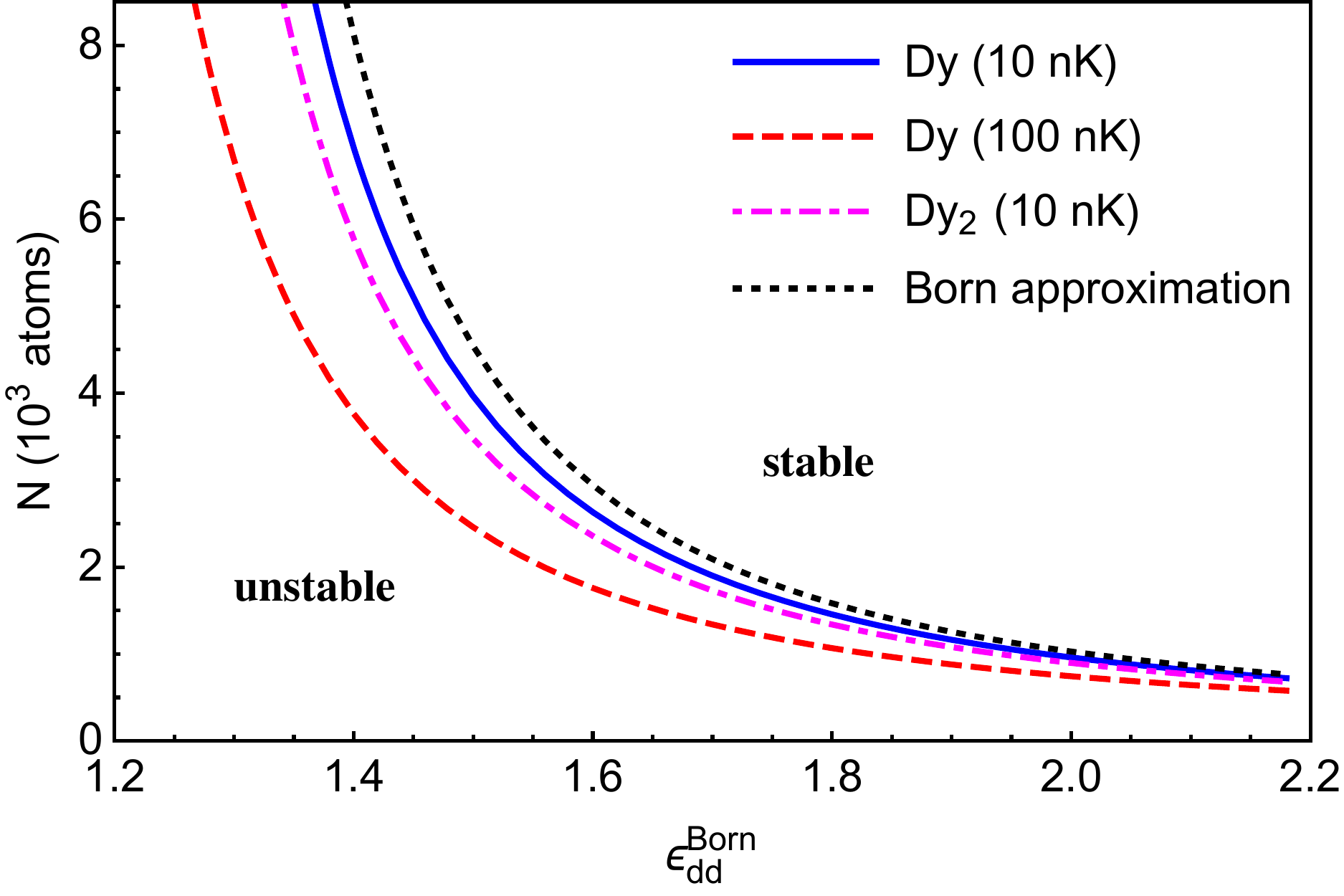}

\caption{\label{fig:selfDy}Droplet stability diagram for Dy and Dy$_2$ with an appropriate $\tilde{a}_{\rm dd}$ as a function of $N$ and $\epsilon_{\rm dd}^{\rm Born}$ calculated using the simplified Gaussian ansatz. The lines correspond to the release energy $E_{\mrm R}=0$ for Dy at 10nK (blue straight line) and 100nK (red dashed line), Dy$_2$ (dot-dashed magenta line) and Born solution (black dotted line). Regions below each line indicates an unstable solution ($E_{\mrm R}>0$), whereas these above mark a stable self-bound state ($E_{\mrm R}<0$). }
\end{figure}

In Fig. \eqref{fig:selfDy} we show the stability diagram for Dy and Dy$_2$ with effective dipolar lengths $a_{\rm dd}$ originated from our model potential compared to the phase boundary obtained within the Born approximation. As shown previously, the self-bound solution stability only depend on the atom number $N$ and $\epsilon_{\rm dd}$~\citep{Baillie2016}. We determine the critical atom number as the value of $N$ at which the release energy $E_{\mrm R}=0$. We plot $N_{\rm crit}$ as a function of $\epsilon_{\rm dd}^{\rm Born}$. The areas above each line correspond to stable (self-bound) solutions with $E_{\rm R}<0$ for a given system, while the areas below each line mark unstable (trap-bound) solutions with $E_{\rm R}>0$. We notice that the qualitative character of the phase diagram is the same for all of the cases, but the lines are offset relative to each other. This is obviously caused by the fact that $\epsilon_{\rm dd} \neq \epsilon_{\rm dd}^{\rm Born}$ for a given scattering length $a$. In the experimentally most interesting regime $\epsilon_{\rm dd}\gtrsim 1$, $\epsilon_{\rm dd} > \epsilon_{\rm dd}^{\rm Born}$. The drop of $\epsilon_{\rm dd}$ at large $a$ is not as relevant since in that regime the droplets do not exist. Note that for Dy$_2$ the critical line is more strongly displaced, which is in line with our findings from Sec. II and Fig. \eqref{fig:add}. These results make the observed phase boundary~\cite{Schmitt2016} more consistent with the previous measurement of the background scattering length~\cite{Maier2015}, as the measured critical atom numbers were also much lower than Born approximation predictions.

\subsection{Collective modes}
\begin{figure}
\centering
\includegraphics[width=0.5\textwidth]{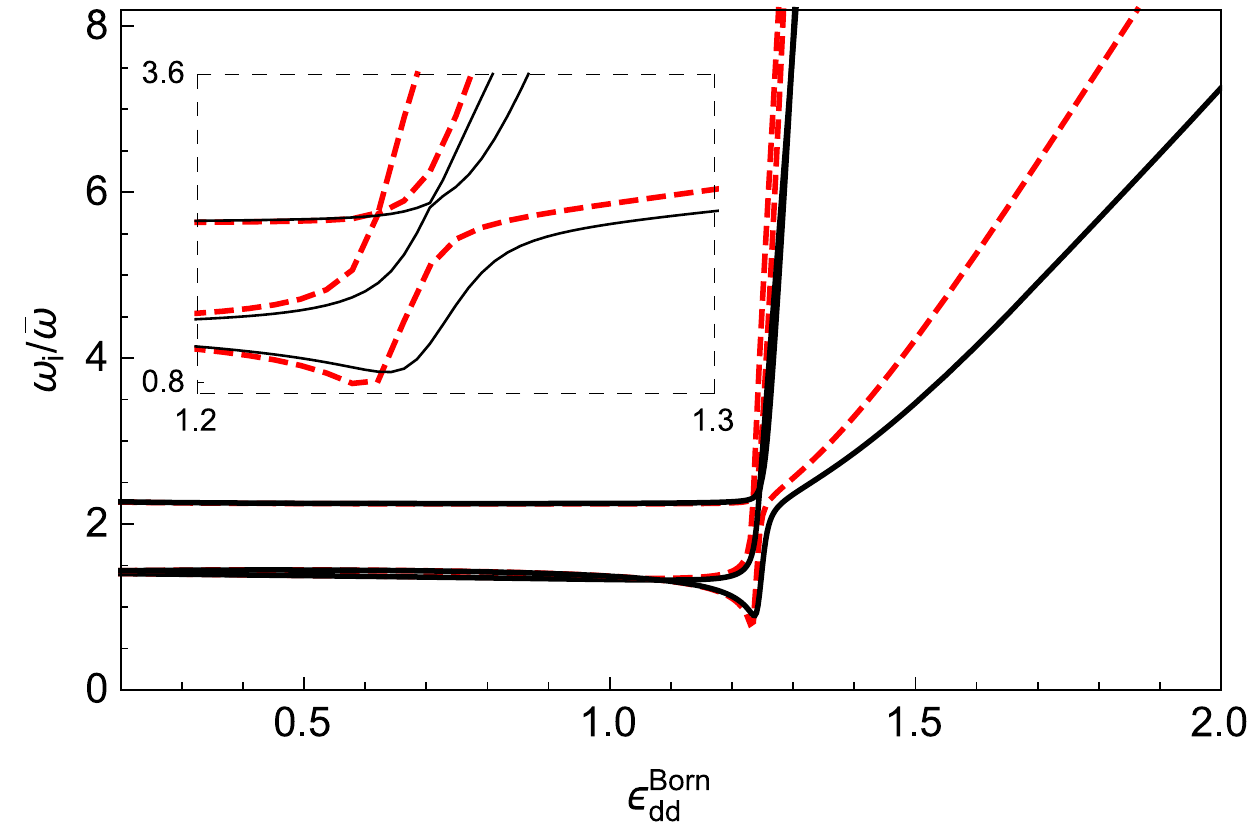}\\
\includegraphics[width=0.5\textwidth]{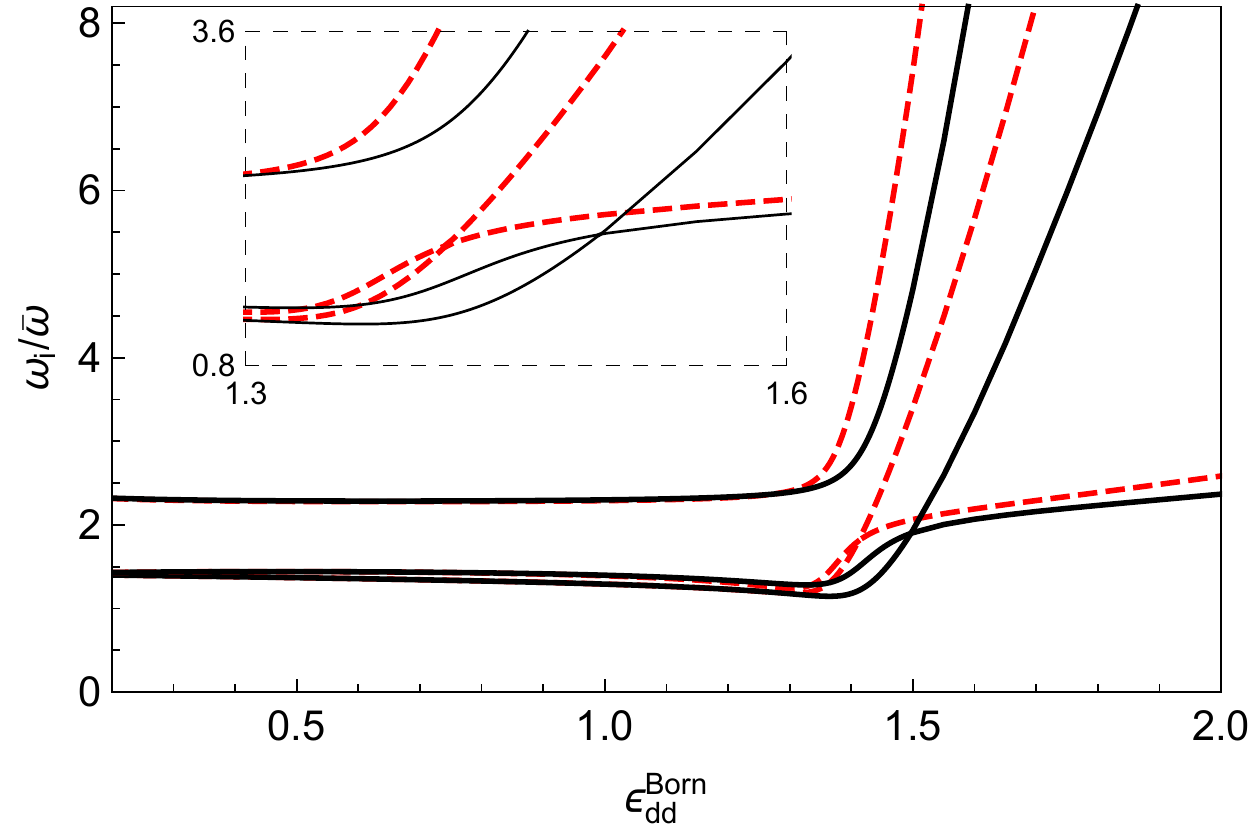}\\
\caption{\label{fig:modesDy}Comparing the lowest excitation frequencies of a spherically trapped Dy BEC with $N=20000$ atoms and $\tilde{\omega}/2\pi=10$ Hz (upper) or $\tilde{\omega}/2\pi=70$ Hz (lower) for the numerical calculation  (black straight line) and Born approximation predictions (red dashed line). In both cases the Gaussian ansatz was used. Both black and red lines exhibit anticrossings. }
\end{figure}

\begin{figure}
\centering
\includegraphics[width=0.5\textwidth]{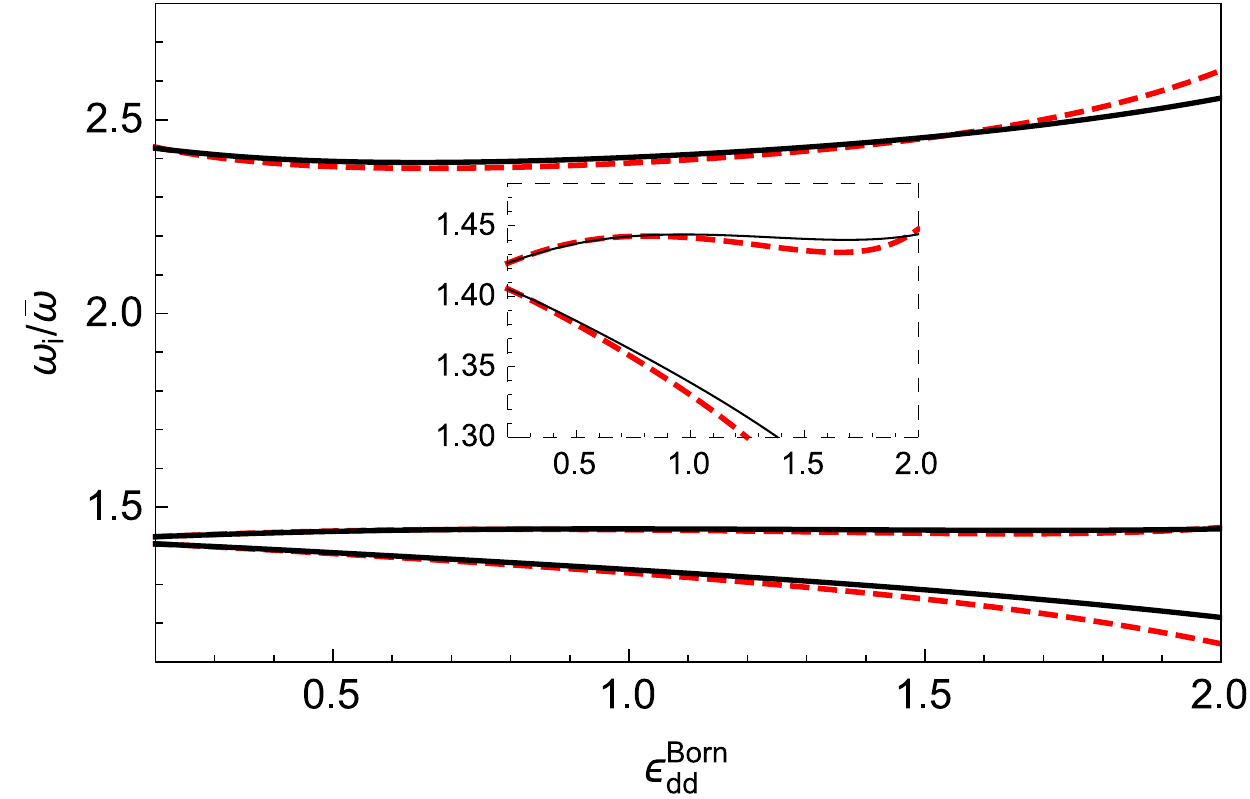}\\
\includegraphics[width=0.5\textwidth]{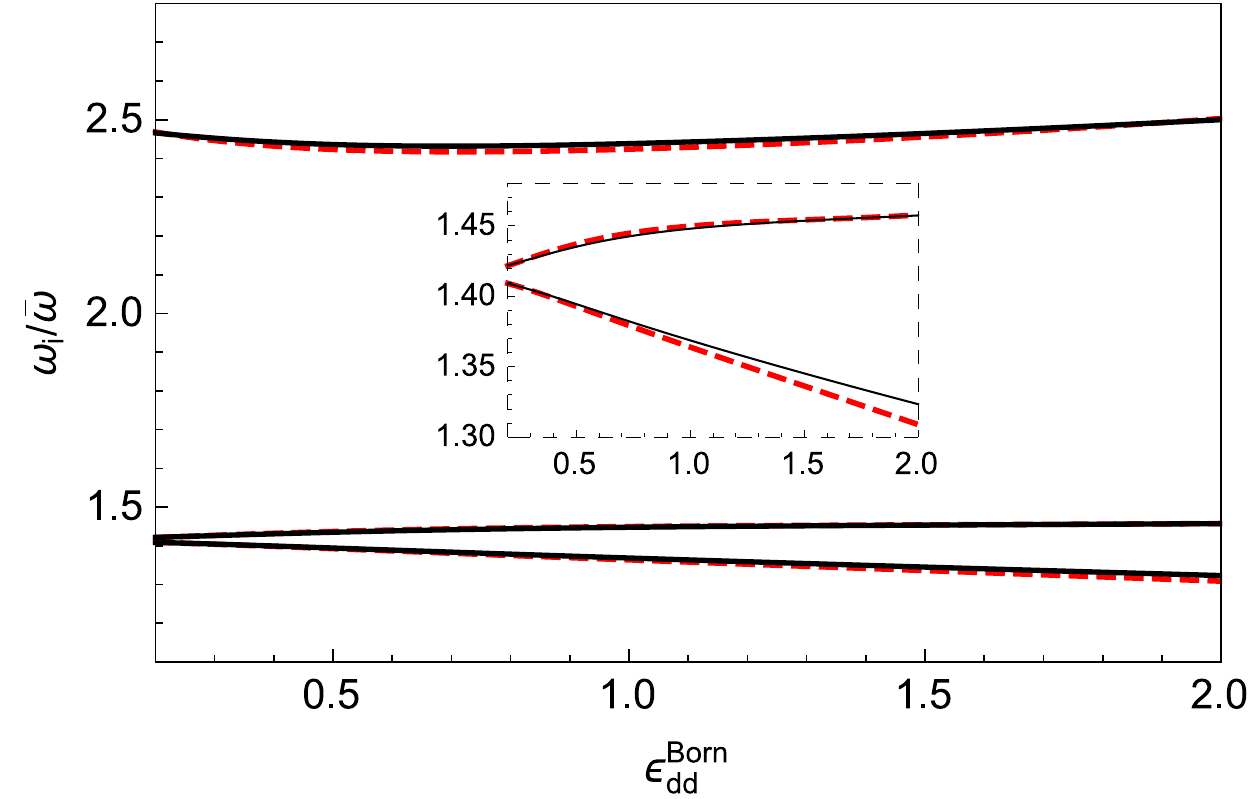}
\caption{\label{fig:modesDy2}Comparing the lowest excitation frequencies of a spherically trapped Dy$_2$ BEC with $N=20000$ atoms and $\tilde{\omega}/2\pi=10$ Hz (upper) or $\tilde{\omega}/2\pi=70$ Hz (lower) for the numerical calculation (black straight line) and Born approximation predictions (red dashed line). In both cases the Gaussian ansatz was used.}
\end{figure}

In order to evaluate the three lowest lying excitations one has to calculate the Hessian matrix for the potential $U$ at the equilibrium widths $\nu_\alpha^{eq}$. In Fig. \eqref{fig:modesDy} and Fig. \eqref{fig:modesDy2} we depict the mode frequencies for Dy and Dy$_2$ respectively, both for the Born approximation and the corrected effective potential. We choose the case of a spherically symmetric trap with $N=$ 20000 atoms for two different trap frequencies $\omega=10$ and $70$ Hz.

In Fig. \eqref{fig:modesDy} we present our findings for Dy. The results for the Born approximation and our model are slightly displaced for both trapping frequencies. The reason for that is the same as in the previous subsection. The composition of the eigenstates is the same as in Ref.~\citep{Wachtler2016a}. Namely, in the mean-field regime ($\epsilon_{\rm dd}^{\rm Born}\approx 1$) the lowest excitation is given by the 2D quadrupole mode, then the second lowest mode is a 3D quadrupole mode, and the highest is a monopole mode. On the other hand, in the droplet regime ($\epsilon_{\rm dd}^{\rm Born} \gg 1$) the mode resembling 3D quadrupole mode becomes the lowest, then the second is the 2D quadrupole mode and the highest mode changes its character from 3D monopole behavior to 2D monopole mode behavior.

The results for Dy$_2$ are depicted in Fig. \eqref{fig:modesDy2}. We notice that in this case there is no significant difference between the Born approximation results and our model. The effect is more visible for a very weak trap. The qualitative difference between the results for Dy$_2$ and Dy can be explained by the fact that the collective frequencies mainly depend on the scattering length $a$ while the dipolar contribution to the Hessian matrix is rather small. As the magnetic moment of Dy$_2$ is higher than of Dy, in order to obtain a given $\epsilon_{\rm dd}^{\rm Born}$ value one needs much higher scattering lengths. Then the contact interaction part completely dominates the dipolar contribution. One can conclude that lanthanide atoms have the perfect combination of parameters for which the dipolar interaction significantly affects the collective modes.

\section{Conclusions}
In conclusion, we studied the effects of taking into account the full scattering amplitude instead of relying on the Born approximation for the dynamics of strongly dipolar bosons gases. We demonstrated that already for lanthanide atoms one needs to use realistic interaction models. We showed that the main effect beyond the Born approximation is the emergence of effective dipolar length which replaces the standard one in the effective potential. This has strong impact on the many-body properties of the gas such as the critical atom number of the self bound droplet and to some extent on collective mode frequencies. Our results suggest that the stability boundary of the self-bound droplets is shifted to smaller atom numbers, in line with the recently reported experimental measurements~\cite{Schmitt2016}. Our findings are relevant also for recently reported Monte Carlo calculations~\cite{Saito2016,Macia2016,Cinti2016} which have used a simplified interaction potential with constant dipolar part and a hard wall, suggesting that more realistic potentials should be used there as well.

So far, the models of dipolar gases were based on local density approximation (LDA). Our treatment is no exception to that, as we calculate the scattering amplitude and the resulting effective interaction and LHY correction in free space. One subtlety in treating dipolar gases is that in order to obtain the scattering amplitude it is necessary to propagate the wave function to distances much larger than $a_{\rm dd}$, comparable to the characteristic trap length. The presence of the trap modifies the asymptotic form of the wave function at this length scale, which can potentially modify the derivation of the mean field theory. Including the trap in the derivation seems to be a challenging task. It would also be desirable to further improve the interaction model, taking into account the anisotropic van der Waals terms~\cite{Petrov:2012} and hyperfine structure details and looking for possible novel effects.

We thank Tomasz Wasak, Jan Kumlin, Hans Peter B\"{u}chler and Igor Ferrier-Barbut for fruitful discussions and comments. RO is grateful to the Stuttgart Dy team for their hospitality.

This work was supported by Alexander von Humboldt Foundation and (Polish) National Science Center Grants No. DEC- 2012/04/A/ST2/00090 and 2014/14/M/ST2/00015.

\bibliography{Allrefs}

\begin{thebibliography}{41}%
\makeatletter
\providecommand \@ifxundefined [1]{%
 \@ifx{#1\undefined}
}%
\providecommand \@ifnum [1]{%
 \ifnum #1\expandafter \@firstoftwo
 \else \expandafter \@secondoftwo
 \fi
}%
\providecommand \@ifx [1]{%
 \ifx #1\expandafter \@firstoftwo
 \else \expandafter \@secondoftwo
 \fi
}%
\providecommand \natexlab [1]{#1}%
\providecommand \enquote  [1]{``#1''}%
\providecommand \bibnamefont  [1]{#1}%
\providecommand \bibfnamefont [1]{#1}%
\providecommand \citenamefont [1]{#1}%
\providecommand \href@noop [0]{\@secondoftwo}%
\providecommand \href [0]{\begingroup \@sanitize@url \@href}%
\providecommand \@href[1]{\@@startlink{#1}\@@href}%
\providecommand \@@href[1]{\endgroup#1\@@endlink}%
\providecommand \@sanitize@url [0]{\catcode `\\12\catcode `\$12\catcode
  `\&12\catcode `\#12\catcode `\^12\catcode `\_12\catcode `\%12\relax}%
\providecommand \@@startlink[1]{}%
\providecommand \@@endlink[0]{}%
\providecommand \url  [0]{\begingroup\@sanitize@url \@url }%
\providecommand \@url [1]{\endgroup\@href {#1}{\urlprefix }}%
\providecommand \urlprefix  [0]{URL }%
\providecommand \Eprint [0]{\href }%
\providecommand \doibase [0]{http://dx.doi.org/}%
\providecommand \selectlanguage [0]{\@gobble}%
\providecommand \bibinfo  [0]{\@secondoftwo}%
\providecommand \bibfield  [0]{\@secondoftwo}%
\providecommand \translation [1]{[#1]}%
\providecommand \BibitemOpen [0]{}%
\providecommand \bibitemStop [0]{}%
\providecommand \bibitemNoStop [0]{.\EOS\space}%
\providecommand \EOS [0]{\spacefactor3000\relax}%
\providecommand \BibitemShut  [1]{\csname bibitem#1\endcsname}%
\let\auto@bib@innerbib\@empty
\bibitem [{\citenamefont {Lahaye}\ \emph {et~al.}(2009)\citenamefont {Lahaye},
  \citenamefont {Menotti}, \citenamefont {Santos}, \citenamefont {Lewenstein},\
  and\ \citenamefont {Pfau}}]{Lahaye:2009}%
  \BibitemOpen
  \bibfield  {author} {\bibinfo {author} {\bibfnamefont {T.}~\bibnamefont
  {Lahaye}}, \bibinfo {author} {\bibfnamefont {C.}~\bibnamefont {Menotti}},
  \bibinfo {author} {\bibfnamefont {L.}~\bibnamefont {Santos}}, \bibinfo
  {author} {\bibfnamefont {M.}~\bibnamefont {Lewenstein}}, \ and\ \bibinfo
  {author} {\bibfnamefont {T.}~\bibnamefont {Pfau}},\ }\href
  {http://stacks.iop.org/0034-4885/72/i=12/a=126401} {\bibfield  {journal}
  {\bibinfo  {journal} {Rep. Prog. Phys.}\ }\textbf {\bibinfo {volume} {72}},\
  \bibinfo {pages} {126401} (\bibinfo {year} {2009})}\BibitemShut {NoStop}%
\bibitem [{\citenamefont {Lahaye}\ \emph {et~al.}(2007)\citenamefont {Lahaye},
  \citenamefont {Koch}, \citenamefont {Fr{\"o}hlich}, \citenamefont {Fattori},
  \citenamefont {Metz}, \citenamefont {Griesmaier}, \citenamefont
  {Giovanazzi},\ and\ \citenamefont {Pfau}}]{Lahaye:2007}%
  \BibitemOpen
  \bibfield  {author} {\bibinfo {author} {\bibfnamefont {T.}~\bibnamefont
  {Lahaye}}, \bibinfo {author} {\bibfnamefont {T.}~\bibnamefont {Koch}},
  \bibinfo {author} {\bibfnamefont {B.}~\bibnamefont {Fr{\"o}hlich}}, \bibinfo
  {author} {\bibfnamefont {M.}~\bibnamefont {Fattori}}, \bibinfo {author}
  {\bibfnamefont {J.}~\bibnamefont {Metz}}, \bibinfo {author} {\bibfnamefont
  {A.}~\bibnamefont {Griesmaier}}, \bibinfo {author} {\bibfnamefont
  {S.}~\bibnamefont {Giovanazzi}}, \ and\ \bibinfo {author} {\bibfnamefont
  {T.}~\bibnamefont {Pfau}},\ }\href {\doibase 10.1038/nature06036} {\bibfield
  {journal} {\bibinfo  {journal} {Nature}\ }\textbf {\bibinfo {volume} {448}},\
  \bibinfo {pages} {672} (\bibinfo {year} {2007})}\BibitemShut {NoStop}%
\bibitem [{\citenamefont {Lu}\ \emph {et~al.}(2011)\citenamefont {Lu},
  \citenamefont {Burdick}, \citenamefont {Youn},\ and\ \citenamefont
  {Lev}}]{Lu:2011}%
  \BibitemOpen
  \bibfield  {author} {\bibinfo {author} {\bibfnamefont {M.}~\bibnamefont
  {Lu}}, \bibinfo {author} {\bibfnamefont {N.~Q.}\ \bibnamefont {Burdick}},
  \bibinfo {author} {\bibfnamefont {S.~H.}\ \bibnamefont {Youn}}, \ and\
  \bibinfo {author} {\bibfnamefont {B.~L.}\ \bibnamefont {Lev}},\ }\href
  {\doibase 10.1103/PhysRevLett.107.190401} {\bibfield  {journal} {\bibinfo
  {journal} {Phys. Rev. Lett.}\ }\textbf {\bibinfo {volume} {107}},\ \bibinfo
  {pages} {190401} (\bibinfo {year} {2011})}\BibitemShut {NoStop}%
\bibitem [{\citenamefont {Aikawa}\ \emph {et~al.}(2012)\citenamefont {Aikawa},
  \citenamefont {Frisch}, \citenamefont {Mark}, \citenamefont {Baier},
  \citenamefont {Rietzler}, \citenamefont {Grimm},\ and\ \citenamefont {{F.
  Ferlaino}}}]{Aikawa:2012}%
  \BibitemOpen
  \bibfield  {author} {\bibinfo {author} {\bibfnamefont {K.}~\bibnamefont
  {Aikawa}}, \bibinfo {author} {\bibfnamefont {A.}~\bibnamefont {Frisch}},
  \bibinfo {author} {\bibfnamefont {M.}~\bibnamefont {Mark}}, \bibinfo {author}
  {\bibfnamefont {S.}~\bibnamefont {Baier}}, \bibinfo {author} {\bibfnamefont
  {A.}~\bibnamefont {Rietzler}}, \bibinfo {author} {\bibfnamefont
  {R.}~\bibnamefont {Grimm}}, \ and\ \bibinfo {author} {\bibnamefont {{F.
  Ferlaino}}},\ }\href {\doibase 10.1103/PhysRevLett.108.210401} {\bibfield
  {journal} {\bibinfo  {journal} {Phys. Rev. Lett.}\ }\textbf {\bibinfo
  {volume} {108}},\ \bibinfo {pages} {210401} (\bibinfo {year}
  {2012})}\BibitemShut {NoStop}%
\bibitem [{\citenamefont {{Ni, K.-K.}}\ \emph {et~al.}(2008)\citenamefont {{Ni,
  K.-K.}}, \citenamefont {Ospelkaus}, \citenamefont {de~Miranda}, \citenamefont
  {Pe'er}, \citenamefont {Neyenhuis}, \citenamefont {Zirbel}, \citenamefont
  {Kotochigova}, \citenamefont {Julienne}, \citenamefont {Jin},\ and\
  \citenamefont {Ye}}]{Ni:2008}%
  \BibitemOpen
  \bibfield  {author} {\bibinfo {author} {\bibnamefont {{Ni, K.-K.}}}, \bibinfo
  {author} {\bibfnamefont {S.}~\bibnamefont {Ospelkaus}}, \bibinfo {author}
  {\bibfnamefont {M.~H.~G.}\ \bibnamefont {de~Miranda}}, \bibinfo {author}
  {\bibfnamefont {A.}~\bibnamefont {Pe'er}}, \bibinfo {author} {\bibfnamefont
  {B.}~\bibnamefont {Neyenhuis}}, \bibinfo {author} {\bibfnamefont {J.~J.}\
  \bibnamefont {Zirbel}}, \bibinfo {author} {\bibfnamefont {S.}~\bibnamefont
  {Kotochigova}}, \bibinfo {author} {\bibfnamefont {P.~S.}\ \bibnamefont
  {Julienne}}, \bibinfo {author} {\bibfnamefont {D.~S.}\ \bibnamefont {Jin}}, \
  and\ \bibinfo {author} {\bibfnamefont {J.}~\bibnamefont {Ye}},\ }\href
  {\doibase 10.1126/science.1163861} {\bibfield  {journal} {\bibinfo  {journal}
  {Science}\ }\textbf {\bibinfo {volume} {322}},\ \bibinfo {pages} {231}
  (\bibinfo {year} {2008})}\BibitemShut {NoStop}%
\bibitem [{\citenamefont {Park}\ \emph {et~al.}(2015)\citenamefont {Park},
  \citenamefont {Will},\ and\ \citenamefont {Zwierlein}}]{Park2015}%
  \BibitemOpen
  \bibfield  {author} {\bibinfo {author} {\bibfnamefont {J.~W.}\ \bibnamefont
  {Park}}, \bibinfo {author} {\bibfnamefont {S.~A.}\ \bibnamefont {Will}}, \
  and\ \bibinfo {author} {\bibfnamefont {M.~W.}\ \bibnamefont {Zwierlein}},\
  }\href@noop {} {\bibfield  {journal} {\bibinfo  {journal} {Physical. Rev.
  Lett.}\ }\textbf {\bibinfo {volume} {114}},\ \bibinfo {pages} {205302}
  (\bibinfo {year} {2015})}\BibitemShut {NoStop}%
\bibitem [{\citenamefont {Wang}\ \emph {et~al.}(2015)\citenamefont {Wang},
  \citenamefont {Julienne},\ and\ \citenamefont {Greene}}]{Wang2015}%
  \BibitemOpen
  \bibfield  {author} {\bibinfo {author} {\bibfnamefont {Y.}~\bibnamefont
  {Wang}}, \bibinfo {author} {\bibfnamefont {P.}~\bibnamefont {Julienne}}, \
  and\ \bibinfo {author} {\bibfnamefont {C.~H.}\ \bibnamefont {Greene}},\
  }\href@noop {} {\bibfield  {journal} {\bibinfo  {journal} {Annual Review of
  Cold Atoms and Molecules}\ }\textbf {\bibinfo {volume} {3}},\ \bibinfo
  {pages} {77} (\bibinfo {year} {2015})}\BibitemShut {NoStop}%
\bibitem [{\citenamefont {Baranov}\ \emph {et~al.}(2012)\citenamefont
  {Baranov}, \citenamefont {Dalmonte}, \citenamefont {Pupillo},\ and\
  \citenamefont {Zoller}}]{baranov2012condensed}%
  \BibitemOpen
  \bibfield  {author} {\bibinfo {author} {\bibfnamefont {M.}~\bibnamefont
  {Baranov}}, \bibinfo {author} {\bibfnamefont {M.}~\bibnamefont {Dalmonte}},
  \bibinfo {author} {\bibfnamefont {G.}~\bibnamefont {Pupillo}}, \ and\
  \bibinfo {author} {\bibfnamefont {P.}~\bibnamefont {Zoller}},\ }\href@noop {}
  {\bibfield  {journal} {\bibinfo  {journal} {Chem. Rev.}\ }\textbf {\bibinfo
  {volume} {112}},\ \bibinfo {pages} {5012} (\bibinfo {year}
  {2012})}\BibitemShut {NoStop}%
\bibitem [{\citenamefont {Lahaye}\ \emph {et~al.}(2008)\citenamefont {Lahaye},
  \citenamefont {Metz}, \citenamefont {Fr\"ohlich}, \citenamefont {Koch},
  \citenamefont {Meister}, \citenamefont {Griesmaier}, \citenamefont {Pfau},
  \citenamefont {Saito}, \citenamefont {Kawaguchi},\ and\ \citenamefont
  {Ueda}}]{Lahaye2008}%
  \BibitemOpen
  \bibfield  {author} {\bibinfo {author} {\bibfnamefont {T.}~\bibnamefont
  {Lahaye}}, \bibinfo {author} {\bibfnamefont {J.}~\bibnamefont {Metz}},
  \bibinfo {author} {\bibfnamefont {B.}~\bibnamefont {Fr\"ohlich}}, \bibinfo
  {author} {\bibfnamefont {T.}~\bibnamefont {Koch}}, \bibinfo {author}
  {\bibfnamefont {M.}~\bibnamefont {Meister}}, \bibinfo {author} {\bibfnamefont
  {A.}~\bibnamefont {Griesmaier}}, \bibinfo {author} {\bibfnamefont
  {T.}~\bibnamefont {Pfau}}, \bibinfo {author} {\bibfnamefont {H.}~\bibnamefont
  {Saito}}, \bibinfo {author} {\bibfnamefont {Y.}~\bibnamefont {Kawaguchi}}, \
  and\ \bibinfo {author} {\bibfnamefont {M.}~\bibnamefont {Ueda}},\ }\href@noop
  {} {\bibfield  {journal} {\bibinfo  {journal} {Phys. Rev. Lett.}\ }\textbf
  {\bibinfo {volume} {101}},\ \bibinfo {pages} {080401} (\bibinfo {year}
  {2008})}\BibitemShut {NoStop}%
\bibitem [{\citenamefont {Ferrier-Barbut}\ \emph {et~al.}(2016)\citenamefont
  {Ferrier-Barbut}, \citenamefont {Kadau}, \citenamefont {Schmitt},
  \citenamefont {Wenzel},\ and\ \citenamefont {Pfau}}]{Barbut2016}%
  \BibitemOpen
  \bibfield  {author} {\bibinfo {author} {\bibfnamefont {I.}~\bibnamefont
  {Ferrier-Barbut}}, \bibinfo {author} {\bibfnamefont {H.}~\bibnamefont
  {Kadau}}, \bibinfo {author} {\bibfnamefont {M.}~\bibnamefont {Schmitt}},
  \bibinfo {author} {\bibfnamefont {M.}~\bibnamefont {Wenzel}}, \ and\ \bibinfo
  {author} {\bibfnamefont {T.}~\bibnamefont {Pfau}},\ }\href {\doibase
  10.1103/PhysRevLett.116.215301} {\bibfield  {journal} {\bibinfo  {journal}
  {Phys. Rev. Lett.}\ }\textbf {\bibinfo {volume} {116}},\ \bibinfo {pages}
  {215301} (\bibinfo {year} {2016})}\BibitemShut {NoStop}%
\bibitem [{\citenamefont {W\"achtler}\ and\ \citenamefont
  {Santos}(2016{\natexlab{a}})}]{Wachtler2016}%
  \BibitemOpen
  \bibfield  {author} {\bibinfo {author} {\bibfnamefont {F.}~\bibnamefont
  {W\"achtler}}\ and\ \bibinfo {author} {\bibfnamefont {L.}~\bibnamefont
  {Santos}},\ }\href {\doibase 10.1103/PhysRevA.93.061603} {\bibfield
  {journal} {\bibinfo  {journal} {Phys. Rev. A}\ }\textbf {\bibinfo {volume}
  {93}},\ \bibinfo {pages} {061603} (\bibinfo {year}
  {2016}{\natexlab{a}})}\BibitemShut {NoStop}%
\bibitem [{\citenamefont {Kadau}\ \emph {et~al.}(2016)\citenamefont {Kadau},
  \citenamefont {Schmitt}, \citenamefont {Wenzel}, \citenamefont {Wink},
  \citenamefont {Maier}, \citenamefont {Ferrier-Barbut},\ and\ \citenamefont
  {Pfau}}]{Kadau2016}%
  \BibitemOpen
  \bibfield  {author} {\bibinfo {author} {\bibfnamefont {H.}~\bibnamefont
  {Kadau}}, \bibinfo {author} {\bibfnamefont {M.}~\bibnamefont {Schmitt}},
  \bibinfo {author} {\bibfnamefont {M.}~\bibnamefont {Wenzel}}, \bibinfo
  {author} {\bibfnamefont {C.}~\bibnamefont {Wink}}, \bibinfo {author}
  {\bibfnamefont {T.}~\bibnamefont {Maier}}, \bibinfo {author} {\bibfnamefont
  {I.}~\bibnamefont {Ferrier-Barbut}}, \ and\ \bibinfo {author} {\bibfnamefont
  {T.}~\bibnamefont {Pfau}},\ }\href@noop {} {\bibfield  {journal} {\bibinfo
  {journal} {Nature}\ }\textbf {\bibinfo {volume} {530}},\ \bibinfo {pages}
  {194} (\bibinfo {year} {2016})}\BibitemShut {NoStop}%
\bibitem [{\citenamefont {W\"achtler}\ and\ \citenamefont
  {Santos}(2016{\natexlab{b}})}]{Wachtler2016a}%
  \BibitemOpen
  \bibfield  {author} {\bibinfo {author} {\bibfnamefont {F.}~\bibnamefont
  {W\"achtler}}\ and\ \bibinfo {author} {\bibfnamefont {L.}~\bibnamefont
  {Santos}},\ }\href {\doibase 10.1103/PhysRevA.94.043618} {\bibfield
  {journal} {\bibinfo  {journal} {Phys. Rev. A}\ }\textbf {\bibinfo {volume}
  {94}},\ \bibinfo {pages} {043618} (\bibinfo {year}
  {2016}{\natexlab{b}})}\BibitemShut {NoStop}%
\bibitem [{\citenamefont {Chomaz}\ \emph {et~al.}(2016)\citenamefont {Chomaz},
  \citenamefont {Baier}, \citenamefont {Petter}, \citenamefont {Mark},
  \citenamefont {W\"achtler}, \citenamefont {Santos},\ and\ \citenamefont
  {Ferlaino}}]{Chomaz2016}%
  \BibitemOpen
  \bibfield  {author} {\bibinfo {author} {\bibfnamefont {L.}~\bibnamefont
  {Chomaz}}, \bibinfo {author} {\bibfnamefont {S.}~\bibnamefont {Baier}},
  \bibinfo {author} {\bibfnamefont {D.}~\bibnamefont {Petter}}, \bibinfo
  {author} {\bibfnamefont {M.~J.}\ \bibnamefont {Mark}}, \bibinfo {author}
  {\bibfnamefont {F.}~\bibnamefont {W\"achtler}}, \bibinfo {author}
  {\bibfnamefont {L.}~\bibnamefont {Santos}}, \ and\ \bibinfo {author}
  {\bibfnamefont {F.}~\bibnamefont {Ferlaino}},\ }\href@noop {} {\bibfield
  {journal} {\bibinfo  {journal} {Phys. Rev. X}\ }\textbf {\bibinfo {volume}
  {6}},\ \bibinfo {pages} {041039} (\bibinfo {year} {2016})}\BibitemShut
  {NoStop}%
\bibitem [{\citenamefont {Baillie}\ \emph {et~al.}(2016)\citenamefont
  {Baillie}, \citenamefont {Wilson}, \citenamefont {Bisset},\ and\
  \citenamefont {Blakie}}]{Baillie2016}%
  \BibitemOpen
  \bibfield  {author} {\bibinfo {author} {\bibfnamefont {D.}~\bibnamefont
  {Baillie}}, \bibinfo {author} {\bibfnamefont {R.~M.}\ \bibnamefont {Wilson}},
  \bibinfo {author} {\bibfnamefont {R.~N.}\ \bibnamefont {Bisset}}, \ and\
  \bibinfo {author} {\bibfnamefont {P.~B.}\ \bibnamefont {Blakie}},\ }\href
  {\doibase 10.1103/PhysRevA.94.021602} {\bibfield  {journal} {\bibinfo
  {journal} {Phys. Rev. A}\ }\textbf {\bibinfo {volume} {94}},\ \bibinfo
  {pages} {021602} (\bibinfo {year} {2016})}\BibitemShut {NoStop}%
\bibitem [{\citenamefont {Schmitt}\ \emph {et~al.}(2016)\citenamefont
  {Schmitt}, \citenamefont {Wenzel}, \citenamefont {B{\"o}ttcher},
  \citenamefont {Ferrier-Barbut},\ and\ \citenamefont {Pfau}}]{Schmitt2016}%
  \BibitemOpen
  \bibfield  {author} {\bibinfo {author} {\bibfnamefont {M.}~\bibnamefont
  {Schmitt}}, \bibinfo {author} {\bibfnamefont {M.}~\bibnamefont {Wenzel}},
  \bibinfo {author} {\bibfnamefont {F.}~\bibnamefont {B{\"o}ttcher}}, \bibinfo
  {author} {\bibfnamefont {I.}~\bibnamefont {Ferrier-Barbut}}, \ and\ \bibinfo
  {author} {\bibfnamefont {T.}~\bibnamefont {Pfau}},\ }\href@noop {} {\bibfield
   {journal} {\bibinfo  {journal} {Nature}\ }\textbf {\bibinfo {volume}
  {539}},\ \bibinfo {pages} {259} (\bibinfo {year} {2016})}\BibitemShut
  {NoStop}%
\bibitem [{\citenamefont {Yi}\ and\ \citenamefont {You}(2001)}]{Yi2001}%
  \BibitemOpen
  \bibfield  {author} {\bibinfo {author} {\bibfnamefont {S.}~\bibnamefont
  {Yi}}\ and\ \bibinfo {author} {\bibfnamefont {L.}~\bibnamefont {You}},\
  }\href {\doibase 10.1103/PhysRevA.63.053607} {\bibfield  {journal} {\bibinfo
  {journal} {Phys. Rev. A}\ }\textbf {\bibinfo {volume} {63}},\ \bibinfo
  {pages} {053607} (\bibinfo {year} {2001})}\BibitemShut {NoStop}%
\bibitem [{\citenamefont {Sch{\"u}tzhold}\ \emph {et~al.}(2006)\citenamefont
  {Sch{\"u}tzhold}, \citenamefont {Uhlmann}, \citenamefont {Xu},\ and\
  \citenamefont {Fischer}}]{Schutzhold2006}%
  \BibitemOpen
  \bibfield  {author} {\bibinfo {author} {\bibfnamefont {R.}~\bibnamefont
  {Sch{\"u}tzhold}}, \bibinfo {author} {\bibfnamefont {M.}~\bibnamefont
  {Uhlmann}}, \bibinfo {author} {\bibfnamefont {Y.}~\bibnamefont {Xu}}, \ and\
  \bibinfo {author} {\bibfnamefont {U.~R.}\ \bibnamefont {Fischer}},\
  }\href@noop {} {\bibfield  {journal} {\bibinfo  {journal} {International
  Journal of Modern Physics B}\ }\textbf {\bibinfo {volume} {20}},\ \bibinfo
  {pages} {3555} (\bibinfo {year} {2006})}\BibitemShut {NoStop}%
\bibitem [{\citenamefont {Lima}\ and\ \citenamefont
  {Pelster}(2011)}]{Lima2011}%
  \BibitemOpen
  \bibfield  {author} {\bibinfo {author} {\bibfnamefont {A.~R.~P.}\
  \bibnamefont {Lima}}\ and\ \bibinfo {author} {\bibfnamefont {A.}~\bibnamefont
  {Pelster}},\ }\href {\doibase 10.1103/PhysRevA.84.041604} {\bibfield
  {journal} {\bibinfo  {journal} {Phys. Rev. A}\ }\textbf {\bibinfo {volume}
  {84}},\ \bibinfo {pages} {041604} (\bibinfo {year} {2011})}\BibitemShut
  {NoStop}%
\bibitem [{\citenamefont {Lima}\ and\ \citenamefont
  {Pelster}(2012)}]{Lima2012}%
  \BibitemOpen
  \bibfield  {author} {\bibinfo {author} {\bibfnamefont {A.~R.~P.}\
  \bibnamefont {Lima}}\ and\ \bibinfo {author} {\bibfnamefont {A.}~\bibnamefont
  {Pelster}},\ }\href {\doibase 10.1103/PhysRevA.86.063609} {\bibfield
  {journal} {\bibinfo  {journal} {Phys. Rev. A}\ }\textbf {\bibinfo {volume}
  {86}},\ \bibinfo {pages} {063609} (\bibinfo {year} {2012})}\BibitemShut
  {NoStop}%
\bibitem [{\citenamefont {Beliaev}(1958)}]{Beliaev1958}%
  \BibitemOpen
  \bibfield  {author} {\bibinfo {author} {\bibfnamefont {S.}~\bibnamefont
  {Beliaev}},\ }\href@noop {} {\bibfield  {journal} {\bibinfo  {journal} {Sov.
  Phys. JETP}\ }\textbf {\bibinfo {volume} {34}},\ \bibinfo {pages} {299}
  (\bibinfo {year} {1958})}\BibitemShut {NoStop}%
\bibitem [{\citenamefont {Hugenholtz}\ and\ \citenamefont
  {Pines}(1959)}]{Hugenholtz1959}%
  \BibitemOpen
  \bibfield  {author} {\bibinfo {author} {\bibfnamefont {N.}~\bibnamefont
  {Hugenholtz}}\ and\ \bibinfo {author} {\bibfnamefont {D.}~\bibnamefont
  {Pines}},\ }\href@noop {} {\bibfield  {journal} {\bibinfo  {journal}
  {Physical Review}\ }\textbf {\bibinfo {volume} {116}},\ \bibinfo {pages}
  {489} (\bibinfo {year} {1959})}\BibitemShut {NoStop}%
\bibitem [{\citenamefont {Kanjilal}\ and\ \citenamefont
  {Blume}(2008)}]{Kanjilal2008}%
  \BibitemOpen
  \bibfield  {author} {\bibinfo {author} {\bibfnamefont {K.}~\bibnamefont
  {Kanjilal}}\ and\ \bibinfo {author} {\bibfnamefont {D.}~\bibnamefont
  {Blume}},\ }\href {\doibase 10.1103/PhysRevA.78.040703} {\bibfield  {journal}
  {\bibinfo  {journal} {Phys. Rev. A}\ }\textbf {\bibinfo {volume} {78}},\
  \bibinfo {pages} {040703} (\bibinfo {year} {2008})}\BibitemShut {NoStop}%
\bibitem [{\citenamefont {Ticknor}(2008)}]{Ticknor2008}%
  \BibitemOpen
  \bibfield  {author} {\bibinfo {author} {\bibfnamefont {C.}~\bibnamefont
  {Ticknor}},\ }\href@noop {} {\bibfield  {journal} {\bibinfo  {journal} {Phys.
  Rev. Lett.}\ }\textbf {\bibinfo {volume} {100}},\ \bibinfo {pages} {133202}
  (\bibinfo {year} {2008})}\BibitemShut {NoStop}%
\bibitem [{\citenamefont {Bohn}\ \emph {et~al.}(2009)\citenamefont {Bohn},
  \citenamefont {Cavagnero},\ and\ \citenamefont {Ticknor}}]{Bohn:2009}%
  \BibitemOpen
  \bibfield  {author} {\bibinfo {author} {\bibfnamefont {J.~L.}\ \bibnamefont
  {Bohn}}, \bibinfo {author} {\bibfnamefont {M.}~\bibnamefont {Cavagnero}}, \
  and\ \bibinfo {author} {\bibfnamefont {C.}~\bibnamefont {Ticknor}},\ }\href
  {\doibase 10.1088/1367-2630/11/5/055039} {\bibfield  {journal} {\bibinfo
  {journal} {New J. Phys.}\ }\textbf {\bibinfo {volume} {11}},\ \bibinfo
  {pages} {055039} (\bibinfo {year} {2009})}\BibitemShut {NoStop}%
\bibitem [{\citenamefont {Zhang}\ and\ \citenamefont {Jie}(2014)}]{Zhang2014}%
  \BibitemOpen
  \bibfield  {author} {\bibinfo {author} {\bibfnamefont {P.}~\bibnamefont
  {Zhang}}\ and\ \bibinfo {author} {\bibfnamefont {J.}~\bibnamefont {Jie}},\
  }\href@noop {} {\bibfield  {journal} {\bibinfo  {journal} {Phys. Rev. A}\
  }\textbf {\bibinfo {volume} {90}},\ \bibinfo {pages} {062714} (\bibinfo
  {year} {2014})}\BibitemShut {NoStop}%
\bibitem [{\citenamefont {Bohn}\ and\ \citenamefont {Jin}(2014)}]{Bohn2014}%
  \BibitemOpen
  \bibfield  {author} {\bibinfo {author} {\bibfnamefont {J.~L.}\ \bibnamefont
  {Bohn}}\ and\ \bibinfo {author} {\bibfnamefont {D.~S.}\ \bibnamefont {Jin}},\
  }\href {\doibase 10.1103/PhysRevA.89.022702} {\bibfield  {journal} {\bibinfo
  {journal} {Phys. Rev. A}\ }\textbf {\bibinfo {volume} {89}},\ \bibinfo
  {pages} {022702} (\bibinfo {year} {2014})}\BibitemShut {NoStop}%
\bibitem [{\citenamefont {Chin}\ \emph {et~al.}(2010)\citenamefont {Chin},
  \citenamefont {Grimm}, \citenamefont {Julienne},\ and\ \citenamefont
  {Tiesinga}}]{Chin:2008}%
  \BibitemOpen
  \bibfield  {author} {\bibinfo {author} {\bibfnamefont {C.}~\bibnamefont
  {Chin}}, \bibinfo {author} {\bibfnamefont {R.}~\bibnamefont {Grimm}},
  \bibinfo {author} {\bibfnamefont {P.}~\bibnamefont {Julienne}}, \ and\
  \bibinfo {author} {\bibfnamefont {E.}~\bibnamefont {Tiesinga}},\ }\href
  {\doibase 10.1103/RevModPhys.82.1225} {\bibfield  {journal} {\bibinfo
  {journal} {Rev. Mod. Phys.}\ }\textbf {\bibinfo {volume} {82}},\ \bibinfo
  {pages} {1225} (\bibinfo {year} {2010})}\BibitemShut {NoStop}%
\bibitem [{\citenamefont {Petrov}\ \emph {et~al.}(2012)\citenamefont {Petrov},
  \citenamefont {Tiesinga},\ and\ \citenamefont {Kotochigova}}]{Petrov:2012}%
  \BibitemOpen
  \bibfield  {author} {\bibinfo {author} {\bibfnamefont {A.}~\bibnamefont
  {Petrov}}, \bibinfo {author} {\bibfnamefont {E.}~\bibnamefont {Tiesinga}}, \
  and\ \bibinfo {author} {\bibfnamefont {S.}~\bibnamefont {Kotochigova}},\
  }\href@noop {} {\bibfield  {journal} {\bibinfo  {journal} {Phys. Rev. Lett.}\
  }\textbf {\bibinfo {volume} {109}},\ \bibinfo {pages} {103002} (\bibinfo
  {year} {2012})}\BibitemShut {NoStop}%
\bibitem [{\citenamefont {Baumann}\ \emph {et~al.}(2014)\citenamefont
  {Baumann}, \citenamefont {Burdick}, \citenamefont {Lu},\ and\ \citenamefont
  {Lev}}]{Baumann:2014}%
  \BibitemOpen
  \bibfield  {author} {\bibinfo {author} {\bibfnamefont {K.}~\bibnamefont
  {Baumann}}, \bibinfo {author} {\bibfnamefont {N.~Q.}\ \bibnamefont
  {Burdick}}, \bibinfo {author} {\bibfnamefont {M.}~\bibnamefont {Lu}}, \ and\
  \bibinfo {author} {\bibfnamefont {B.~L.}\ \bibnamefont {Lev}},\ }\href
  {\doibase 10.1103/PhysRevA.89.020701} {\bibfield  {journal} {\bibinfo
  {journal} {Phys. Rev. A}\ }\textbf {\bibinfo {volume} {89}},\ \bibinfo
  {pages} {020701} (\bibinfo {year} {2014})}\BibitemShut {NoStop}%
\bibitem [{\citenamefont {Frisch}\ \emph {et~al.}(2014)\citenamefont {Frisch},
  \citenamefont {Mark}, \citenamefont {Aikawa}, \citenamefont {Ferlaino},
  \citenamefont {Bohn}, \citenamefont {Makrides}, \citenamefont {Petrov},\ and\
  \citenamefont {Kotochigova}}]{Frisch2014}%
  \BibitemOpen
  \bibfield  {author} {\bibinfo {author} {\bibfnamefont {A.}~\bibnamefont
  {Frisch}}, \bibinfo {author} {\bibfnamefont {M.}~\bibnamefont {Mark}},
  \bibinfo {author} {\bibfnamefont {K.}~\bibnamefont {Aikawa}}, \bibinfo
  {author} {\bibfnamefont {F.}~\bibnamefont {Ferlaino}}, \bibinfo {author}
  {\bibfnamefont {J.~L.}\ \bibnamefont {Bohn}}, \bibinfo {author}
  {\bibfnamefont {C.}~\bibnamefont {Makrides}}, \bibinfo {author}
  {\bibfnamefont {A.}~\bibnamefont {Petrov}}, \ and\ \bibinfo {author}
  {\bibfnamefont {S.}~\bibnamefont {Kotochigova}},\ }\href@noop {} {\bibfield
  {journal} {\bibinfo  {journal} {Nature}\ }\textbf {\bibinfo {volume} {507}},\
  \bibinfo {pages} {475} (\bibinfo {year} {2014})}\BibitemShut {NoStop}%
\bibitem [{\citenamefont {Maier}\ \emph {et~al.}(2015)\citenamefont {Maier},
  \citenamefont {Ferrier-Barbut}, \citenamefont {Kadau}, \citenamefont
  {Schmitt}, \citenamefont {Wenzel}, \citenamefont {Wink}, \citenamefont
  {Pfau}, \citenamefont {Jachymski},\ and\ \citenamefont
  {Julienne}}]{Maier2015}%
  \BibitemOpen
  \bibfield  {author} {\bibinfo {author} {\bibfnamefont {T.}~\bibnamefont
  {Maier}}, \bibinfo {author} {\bibfnamefont {I.}~\bibnamefont
  {Ferrier-Barbut}}, \bibinfo {author} {\bibfnamefont {H.}~\bibnamefont
  {Kadau}}, \bibinfo {author} {\bibfnamefont {M.}~\bibnamefont {Schmitt}},
  \bibinfo {author} {\bibfnamefont {M.}~\bibnamefont {Wenzel}}, \bibinfo
  {author} {\bibfnamefont {C.}~\bibnamefont {Wink}}, \bibinfo {author}
  {\bibfnamefont {T.}~\bibnamefont {Pfau}}, \bibinfo {author} {\bibfnamefont
  {K.}~\bibnamefont {Jachymski}}, \ and\ \bibinfo {author} {\bibfnamefont
  {P.~S.}\ \bibnamefont {Julienne}},\ }\href@noop {} {\bibfield  {journal}
  {\bibinfo  {journal} {Phys. Rev. A}\ }\textbf {\bibinfo {volume} {92}},\
  \bibinfo {pages} {060702(R)} (\bibinfo {year} {2015})}\BibitemShut {NoStop}%
\bibitem [{\citenamefont {Wang}(2008)}]{Wang2008}%
  \BibitemOpen
  \bibfield  {author} {\bibinfo {author} {\bibfnamefont {D.-W.}\ \bibnamefont
  {Wang}},\ }\href@noop {} {\bibfield  {journal} {\bibinfo  {journal} {New J.
  Phys.}\ }\textbf {\bibinfo {volume} {10}},\ \bibinfo {pages} {053005}
  (\bibinfo {year} {2008})}\BibitemShut {NoStop}%
\bibitem [{\citenamefont {Huang}\ \emph {et~al.}(2010)\citenamefont {Huang},
  \citenamefont {Wang},\ and\ \citenamefont {Wu}}]{Huang2010}%
  \BibitemOpen
  \bibfield  {author} {\bibinfo {author} {\bibfnamefont {C.-C.}\ \bibnamefont
  {Huang}}, \bibinfo {author} {\bibfnamefont {D.-W.}\ \bibnamefont {Wang}}, \
  and\ \bibinfo {author} {\bibfnamefont {W.-C.}\ \bibnamefont {Wu}},\
  }\href@noop {} {\bibfield  {journal} {\bibinfo  {journal} {Phys. Rev. A}\
  }\textbf {\bibinfo {volume} {81}},\ \bibinfo {pages} {043629} (\bibinfo
  {year} {2010})}\BibitemShut {NoStop}%
\bibitem [{\citenamefont {Boudjemaa}(2016)}]{Boudjemaa2016}%
  \BibitemOpen
  \bibfield  {author} {\bibinfo {author} {\bibfnamefont {A.}~\bibnamefont
  {Boudjemaa}},\ }\href@noop {} {\bibfield  {journal} {\bibinfo  {journal}
  {arXiv preprint arXiv:1610.00489}\ } (\bibinfo {year} {2016})}\BibitemShut
  {NoStop}%
\bibitem [{\citenamefont {O{\l}dziejewski}\ \emph {et~al.}(2016)\citenamefont
  {O{\l}dziejewski}, \citenamefont {G{\'{o}}recki},\ and\ \citenamefont
  {Rzazewski}}]{Oldziejewski2016}%
  \BibitemOpen
  \bibfield  {author} {\bibinfo {author} {\bibfnamefont {R.}~\bibnamefont
  {O{\l}dziejewski}}, \bibinfo {author} {\bibfnamefont {W.}~\bibnamefont
  {G{\'{o}}recki}}, \ and\ \bibinfo {author} {\bibfnamefont {K.}~\bibnamefont
  Rzazewski},\ }\href {\doibase 10.1209/0295-5075/114/46003}
  {\bibfield  {journal} {\bibinfo  {journal} {EPL}\ }\textbf {\bibinfo {volume}
  {114}},\ \bibinfo {pages} {46003} (\bibinfo {year} {2016})}\BibitemShut
  {NoStop}%
\bibitem [{\citenamefont {Li}\ \emph {et~al.}(2016)\citenamefont {Li},
  \citenamefont {Wyart}, \citenamefont {Dulieu}, \citenamefont {Nascimbene},\
  and\ \citenamefont {Lepers}}]{Lepers2016}%
  \BibitemOpen
  \bibfield  {author} {\bibinfo {author} {\bibfnamefont {H.}~\bibnamefont
  {Li}}, \bibinfo {author} {\bibfnamefont {J.-F.}\ \bibnamefont {Wyart}},
  \bibinfo {author} {\bibfnamefont {O.}~\bibnamefont {Dulieu}}, \bibinfo
  {author} {\bibfnamefont {S.}~\bibnamefont {Nascimbene}}, \ and\ \bibinfo
  {author} {\bibfnamefont {M.}~\bibnamefont {Lepers}},\ }\href@noop {}
  {\bibfield  {journal} {\bibinfo  {journal} {arXiv preprint 1607.05628}\ }
  (\bibinfo {year} {2016})}\BibitemShut {NoStop}%
\bibitem [{\citenamefont {Frisch}\ \emph {et~al.}(2015)\citenamefont {Frisch},
  \citenamefont {Mark}, \citenamefont {Aikawa}, \citenamefont {Baier},
  \citenamefont {Grimm}, \citenamefont {Petrov}, \citenamefont {Kotochigova},
  \citenamefont {Qu{\'e}m{\'e}ner}, \citenamefont {Lepers}, \citenamefont
  {Dulieu},\ and\ \citenamefont {Ferlaino}}]{Frisch:2015}%
  \BibitemOpen
  \bibfield  {author} {\bibinfo {author} {\bibfnamefont {A.}~\bibnamefont
  {Frisch}}, \bibinfo {author} {\bibfnamefont {M.}~\bibnamefont {Mark}},
  \bibinfo {author} {\bibfnamefont {K.}~\bibnamefont {Aikawa}}, \bibinfo
  {author} {\bibfnamefont {S.}~\bibnamefont {Baier}}, \bibinfo {author}
  {\bibfnamefont {R.}~\bibnamefont {Grimm}}, \bibinfo {author} {\bibfnamefont
  {A.}~\bibnamefont {Petrov}}, \bibinfo {author} {\bibfnamefont
  {S.}~\bibnamefont {Kotochigova}}, \bibinfo {author} {\bibfnamefont
  {G.}~\bibnamefont {Qu{\'e}m{\'e}ner}}, \bibinfo {author} {\bibfnamefont
  {M.}~\bibnamefont {Lepers}}, \bibinfo {author} {\bibfnamefont
  {O.}~\bibnamefont {Dulieu}}, \ and\ \bibinfo {author} {\bibfnamefont
  {F.}~\bibnamefont {Ferlaino}},\ }\href@noop {} {\bibfield  {journal}
  {\bibinfo  {journal} {ArXiv.org}\ } (\bibinfo {year} {2015})},\ \Eprint
  {http://arxiv.org/abs/1504.04578} {1504.04578} \BibitemShut {NoStop}%
\bibitem [{\citenamefont {Saito}(2016)}]{Saito2016}%
  \BibitemOpen
  \bibfield  {author} {\bibinfo {author} {\bibfnamefont {H.}~\bibnamefont
  {Saito}},\ }\href@noop {} {\bibfield  {journal} {\bibinfo  {journal} {Journal
  of the Physical Society of Japan}\ }\textbf {\bibinfo {volume} {85}},\
  \bibinfo {pages} {053001} (\bibinfo {year} {2016})}\BibitemShut {NoStop}%
\bibitem [{\citenamefont {Macia}\ \emph {et~al.}(2016)\citenamefont {Macia},
  \citenamefont {S\'anchez-Baena}, \citenamefont {Boronat},\ and\ \citenamefont
  {Mazzanti}}]{Macia2016}%
  \BibitemOpen
  \bibfield  {author} {\bibinfo {author} {\bibfnamefont {A.}~\bibnamefont
  {Macia}}, \bibinfo {author} {\bibfnamefont {J.}~\bibnamefont
  {S\'anchez-Baena}}, \bibinfo {author} {\bibfnamefont {J.}~\bibnamefont
  {Boronat}}, \ and\ \bibinfo {author} {\bibfnamefont {F.}~\bibnamefont
  {Mazzanti}},\ }\href {\doibase 10.1103/PhysRevLett.117.205301} {\bibfield
  {journal} {\bibinfo  {journal} {Phys. Rev. Lett.}\ }\textbf {\bibinfo
  {volume} {117}},\ \bibinfo {pages} {205301} (\bibinfo {year}
  {2016})}\BibitemShut {NoStop}%
\bibitem [{\citenamefont {Cinti}\ \emph {et~al.}(2016)\citenamefont {Cinti},
  \citenamefont {Cappellaro}, \citenamefont {Salasnich},\ and\ \citenamefont
  {Macr{\`\i}}}]{Cinti2016}%
  \BibitemOpen
  \bibfield  {author} {\bibinfo {author} {\bibfnamefont {F.}~\bibnamefont
  {Cinti}}, \bibinfo {author} {\bibfnamefont {A.}~\bibnamefont {Cappellaro}},
  \bibinfo {author} {\bibfnamefont {L.}~\bibnamefont {Salasnich}}, \ and\
  \bibinfo {author} {\bibfnamefont {T.}~\bibnamefont {Macr{\`\i}}},\
  }\href@noop {} {\bibfield  {journal} {\bibinfo  {journal} {arXiv preprint
  arXiv:1610.03119}\ } (\bibinfo {year} {2016})}\BibitemShut {NoStop}%
\end{thebibliography}%
\end{document}